\documentclass[prd,showpacs,showkeys,nofootinbib]{revtex4}
\usepackage{psfrag}
\usepackage{graphicx}
\usepackage{dcolumn}
\usepackage{bm}
\usepackage{subfigure}
\newcommand{\be}{\begin{equation}}
\newcommand{\ee}{\end{equation}}
\newcommand{\bea}{\begin{eqnarray}}
\newcommand{\eea}{\end{eqnarray}}
\newcommand{\ba}{\begin{array}}
\newcommand{\ea}{\end{array}}
\newcommand{\lsim}
{{\;\raise0.3ex\hbox{$<$\kern-0.75em\raise-1.1ex\hbox{$\sim$}}\;}}
\newcommand{\gsim}
{{\;\raise0.3ex\hbox{$>$\kern-0.75em\raise-1.1ex\hbox{$\sim$}}\;}}
\def\mZsqc2b{\tilde M_Z^2 \cos(2\beta)}
\def\mZsqc2bst{\tilde M_Z^2 \cos(2\beta) \sin^2 \theta_W}
\def\msQO{m^2_{\tilde Q_L}}
\def\msQ{M^2_{\tilde Q_L}}

\def\msuL{M^2_{\tilde u_L}}
\def\muL{M_{\tilde u_L}}
\def\muR{M_{\tilde u_R}}
\def\tuL{M_{\tilde u_L}}
\def\msdL{M^2_{\tilde d_L}}
\def\tdl{M_{\tilde d_L}}
\def\msuR{M^2_{\tilde u_R}}
\def\msuRO{m^2_{\tilde u_R}}
\def\tuR{M_{\tilde u_R}}
\def\msdR{M^2_{\tilde d_R}}
\def\mdR{M_{\tilde d_R}}
\def\mdL{M_{\tilde d_L}}
\def\msdRO{m^2_{\tilde d_R}}
\def\tdR{M_{\tilde d_R}}
\def\mslL{M^2_{\tilde L_L}}
\def\mslLO{m^2_{\tilde L_L}}

\def\mseL{M^2_{\tilde e_L}}

\def\teL{M_{\tilde e_L}}
\def\msnL{M^2_{\tilde{\nu}_L}}

\def\tnL{M_{\tilde{\nu}_L}}
\def\mseR{M^2_{\tilde e_R}}
\def\meR{M_{\tilde e_R}}
\def\mseRO{m^2_{\tilde e_R}}
\def\teR{M_{\tilde e_R}}

\def\mhUO{m^2_{H_u}}
\def\mhDO{m^2_{H_d}}
\def\MZsq{M_Z^2}
\def\ctwoB{\cos(2\beta)}
\def\msq{m_0^2}
\def\msixteensq{m_{16}^2}
\def\mtensq{m_{10}^2}
\def\mfivesq{m_{\overline{5}}^2}
\def\stwoW{\sin^2\theta_W}
\def\s2W{\sin^2\theta_W}

\def\cone{\overline{c}_1}
\def\ctwo{\overline{c}_2}
\def\cthree{\overline{c}_3}
\def\sul{\tilde u_L}
\def\sur{\tilde u_R}
\def\sdl{\tilde d_L}
\def\sdr{\tilde d_R}
\def\ssl{\tilde s_L}
\def\ssr{\tilde s_R}
\def\scl{\tilde c_L}
\def\scr{\tilde c_R}

\def\sel{\tilde e_L}
\def\ser{\tilde e_R}
\def\cuL{c_{\tilde{u}_L}}
\def\cdL{c_{\tilde{d}_L}}
\def\cuR{c_{\tilde{u}_R}}
\def\cdR{c_{\tilde{d}_R}}
\def\ceR{c_{\tilde{e}_R}}
\def\smul{\tilde \mu_L}
\def\smur{\tilde \mu_R}

\def\seln{\tilde \nu_e}
\def\smun{\tilde \nu_\mu}

\def\st{\sin^2 \theta_W}

\begin{document}
\bigskip
\title{\bf Sparticle Mass Spectrum in Grand Unified Theories}
\author{B. Ananthanarayan}
\affiliation{Centre for High Energy Physics, Indian Institute of 
Science, Bangalore 560 012, India} 
\author{P. N. Pandita}
\affiliation{ The Institute of Mathematical Sciences, Chennai 600 113, India \\
Department of Physics,
North-Eastern Hill University,  Shillong 793 022, 
India~\footnote{Permanent address}}
\begin{abstract}
We carry out a detailed analysis of sparticle mass spectrum
in supersymmetric grand unified theories. We consider 
the spectroscopy of the squarks and sleptons in $SU(5)$ and $SO(10)$ 
grand unified theories, and show how the underlying supersymmetry
breaking parameters of these theories can be determined from a
measurement of different sparticle masses. This analysis is done 
analytically by integrating the one-loop renormalization group
equations with appropriate boundary conditions implied by the
underlying grand unified gauge group.
We also consider the impact of non-universal gaugino masses
on the sparticle spectrum, especially the neutralino and 
chargino masses which arise in
supersymmetric grand unified theories with  
non-minimal gauge kinetic function. In particular,
we study the interrelationships between the squark and slepton masses
which arise in grand unified theories at the one-loop level, 
which can be used to distinguish between the different underlying gauge groups
and their breaking pattern to the Standard Model gauge group. We also
comment on the corrections that can affect these one-loop results. 
\end{abstract}

\maketitle

\section{\label{sec:intro} Introduction}

Despite its great  success, the gauge group $SU(3)\times
SU(2)\times U(1)$ remains a completely unexplained feature of
the Standard Model~(SM) of electroweak and strong interactions.
The idea of grand unification~\cite{GG1}
is, therefore,  one of the most compelling theoretical
ideas that goes beyond the Standard Model.
In grand unified theories~(GUTs), the SM gauge group
can be elegantly unified into a simple group. Moreover, the fermion
content of the SM can be accomodated in  irreducible
representations of the unified gauge group. Also, one can understand
the smallness of neutrino masses via the  seesaw 
mechanism~\cite{Seesaw1,Seesaw2,Seesaw3,Seesaw4}
in some of the grand unified  models~\cite{SO101,SO102}  like $SO(10)$.
The renormalization group flow of the gauge couplings leads to
their unification at a very large scale~\cite{GQW}. However, this
picture wherein the SM is embedded into a grand unified theory~(GUT) 
with gauge coupling unification
at large scale leads to the well known hierarchy and naturalness
problems due to the widely separated scales,  the weak scale 
characterized by the mass of the $Z$-boson~($\sim M_Z$),
and the large unification scale characterized by the gauge coupling
unification.

Supersymmetry~(SUSY) is at present the only  known framework~\cite{wess}
in which the hierarchy between the weak scale and the large GUT scale
can be made technically natural~\cite{kaul1,kaul2,kaul3,kaul4,kaul5}.
Supersymmetry is, however,  not an exact symmetry in nature.
The precise manner in which SUSY is broken is not known  at present.
However, the necessary SUSY breaking can be introduced through soft
supersymmetry breaking terms
that do not reintroduce quadratic divergences in the Higgs mass,
and thereby do not disturb the stability of the hierarchy between the 
weak scale and the  GUT scale. Such terms can typically 
arise in  supergravity
theories, in which  local supersymmetry is 
spontaneously broken in a hidden sector, and is then transmitted to 
the visible sector via gravitational interactions. 
This is what is usually done 
in the case of the minimal supersymmetric
standard model~(MSSM), 
with gravity mediated supersymmetry 
breaking~\cite{Nilles, LSusyreview}. However, minimality is only a simplifying
assumption, and may not necessairly lead to realistic models.

At present the most direct
phenomenological evidence in favour of supersymmetry is obtained from 
the unification of couplings in GUTs. Precise LEP data on 
$\alpha_s(m_Z)$ and $\sin^2{\theta_W}$ show that
standard one-scale GUTs fail in predicting $\sin^2{\theta_W}$,  given
$\alpha_s(m_Z)$ (and $\alpha(m_Z)$),  while GUTs based on 
supersymmetry~(SUSY GUTs)~\cite{dimopoulos}
are in agreement with the present experimental results. 
If one starts from the known values of $\sin^2{\theta_W}$ and 
$\alpha(m_Z)$, 
one finds \cite{Lang1,Lang2,Lang3} for $\alpha_s(m_Z)$ the results:
$\alpha_s(m_Z) = 0.073\pm 0.002$ for Standard GUTs and 
$\alpha_s(m_Z) = 0.129\pm0.010$ for SUSY GUTs
to be compared with the world average experimental value 
$\alpha_s(m_Z) =0.118\pm0.002$.  Furthermore,  one of the most important 
predictions of grand unification is
that, because of the presence of baryon number violating
interactions, proton must decay. In SUSY GUTs
proton decay is much slower as compared to the non SUSY case. 
This is because the unification mass is typically 
$M_{GUT}\sim~\rm{few}~10^{16}~GeV$ in  SUSY GUTs, which is about 20-30 
times larger than 
for ordinary GUTs. This makes proton  decay via gauge
boson exchange negligible and the main decay amplitude arises from 
dimension-5 operators with higgsino exchange, leading to a
rate close but still compatible with existing bounds~\cite{AFM}. 
Moreover, SUSY provides an excellent dark matter candidate, 
the neutralino.  
We finally recall that the range of neutrino masses as indicated by 
oscillation experiments, when interpreted in the see-saw mechanism, 
point toward a large scale~\cite{alfe} and give additional support 
to GUTs.

This naturally leads us to the idea of supersymmetric grand unification.
In supersymmetric grand unified theories the soft terms which break 
supersymmetry are introduced at some large scale~(the GUT scale), from
where they evolve through renormalization group equations~(RGEs) to the
electroweak scale~\cite{Inoue:1982pi}. The values of the soft terms
so evolved to the weak scale are then used to make predictions for the
masses of the superpartners of the SM particles. However, this 
leaves the question of underlying grand unified gauge group open.
In this paper we consider the question whether the underlying 
grand unified gauge group leaves its imprint on the superpartner 
masses. This question is of great importance as the Large 
Hadron Collider~(LHC) is expected to
start operating within the next couple of years and is likely to discover
the supersymmetric partners of the SM particles. Furthemore,
there is a  possibility of an International Linear Collider~(ILC) 
being constructed, where
a precision study of the properties of these states is likely
to become  a reality~\cite{LHCILCReview}. It is, therefore, 
important to consider what may be learnt from measurement
of superpartner masses about the underlying supersymmetric grand
unified theory.  This question of reconstruction of the 
underlying parameters~\cite{FHKN, BPZ, Blair, Kane}
of a supersymmetric theory from a measurement of superpartner masses 
and then using these parameters to distinguish between 
different underlying grand unified gauge groups
is precisely the one which can be addressed 
in the framework of renormalization group evolution of  these 
parameters. This is because the underlying grand  unified 
gauge group leaves its imprint on the sparticle 
spectrum through the boundary conditions at the grand unified scale
which serve as an input to the RG evolution of these parameters.
Thus, by measuring the masses of the sparticles, and hence
reconstructing
the supersymmetric parameters at the large scale from these masses, 
we can determine the  GUT gauge group of the underlying
supersymmetric grand unified theory, and its breaking pattern
to the SM gauge group. 

In this paper we consider the sparticle spectrum in supersymmetric
grand unified theories in detail in order to
determine the parameters of the underlying theory at the large
GUT scale, which can then be used to determine the 
underlying grand unified gauge group. We point out at the 
outset that this analysis is done 
by analytically integrating the one-loop renormalization group 
equations with boundary conditions appropriate to the grand unified
gauge group. However, we note that two- and three-loop 
contributions to the renormalization group equations may have
significant impact on the results based on the one-loop
renormalization group equations. In addition shifts from the
$\overline{\rm DR}$ to the on-shell scheme, as well as theoretical
uncertainities may also affect the relations between the high-scale
parameters and the physical mass spectrum. Moreover, the uncertainties
in the SUSY breaking scale as well as the experimental uncertainties
on the sparticle mass spectrum will affect the determination of 
the underlying parameters. However, results obtained on the basis of
analytical solutions of the one-loop renormalization, which are
possible only at the one-loop level, are physically transparent,
and can serve as a basis for a more precise  numerical ananlysis. 

We recall that the SM gauge group can be embedded into a larger 
gauge group, where an entire SM generation can be fitted into
a single (ir)reducible representation of the underlying gauge group.
Indeed, there is chain of group embeddings~\cite{Ramond:1979py}
of the SM gauge group into a larger group
$ {SU(3)_c \times SU(2)_L \times U(1)_Y 
\subset SU(5)\subset SO(10)\subset E_6\subset E_7\subset E_8.}$
However, in four-dimensional grand unified theories the gauge groups
$E_7$ and $E_8$ do not support a chiral structure
of the weak interactions, and hence cannot be
used as grand unified gauge groups in four dimensions. 
This leaves out only the three groups,
$SU(5)$, $SO(10)$, and  $E_6$ as possible unified gauge
groups in four dimensions. In this paper we shall consider 
grand unified supersymmetric theories based on the gauge groups
$SU(5)$ and $SO(10).$ 

As pointed above, minimality need not lead to realistic models.
In grand unified theories there can be departures
from the minimality that is assumed in the minimal supersymmetric
standard model. The nonminimality can arise because of the boundary 
conditions at the grand unified scale. This manifests itself in the
form of non-universal soft scalar masses at the scale of the breaking of the
grand unified gauge group. For example, in the $SU(5)$ supersymmetric
grand unified theory, since the fermions~(and the sfermions) 
belong to the $\bf 10 $, and $\bf \bar 5 $ representations, respectively, the
soft scalar masses are different for sfermions belonging to these
representations. On the other hand in the $SO(10)$ unification, although 
all the fermions~(and the sfermions) belong  to a single $\bf 16$-dimensional
representation of the gauge group, the boundary conditions 
for soft scalar masses 
are non-universal because of the $D$-term contributions
to these masses at the GUT scale. These
$D$-term contributions to the soft scalar
masses arise because the rank of $SO(10)$ is higher than SM gauge group. 
In general $D-$term contributions
to the SUSY breaking soft scalar masses arise whenever a gauge symmetry is
spontaneously broken with a reduction of rank~\cite{Drees:1986vd}.
These $D-$term contributions can have important phenomenological
consequences at low energies as they allow one to reach certain 
regions of parameter space which are not otherwise accessible with  
universal boundary 
conditions~\cite{Kolda:1995iw, Auto:2003ys, ChengHall}.
In particular, these $D$-term contributions
are likely to help distinguish between different scenarios for 
breaking of grand unified gauge group to the Standard Model
gauge group~\cite{Kawamura1,Kawamura2}.

Another source of departure from the universality can arise
in the gaugino sector. Gaugino masses arise 
from higher dimensional interaction
terms which involve gauginos and auxiliary parts of chiral superfields
in a given supersymmetric  model.
For example, in the $SU(5)$ supersymmetric grand unified theory the 
auxiliary part of a chiral superfield
in these higher dimensional interaction terms can be in the
representation {\bf 1}, {\bf 24}, {\bf 75}, or {\bf 200}, or some
combination of these, of the $SU(5)$ gauge group. If the
auxiliary field of one of the  $SU(5)$ nonsinglet chiral superfields
obtains a  vacuum expectation value (VEV), then the
gaugino masses are not universal at the grand unification
scale~\cite{EENT,DreesKE}. Since the 
phenomenology~\cite{coll1,coll2,constr1,constr2,DMM,HLPR,KLNPY} 
of supersymmetric models depends
crucially on the composition of neutralinos and charginos, it
is important to investigate the changes in the experimental signals
for supersymmetry with the changes in the composition of neutralinos
and charginos that may arise because of the changes in the underlying
boundary conditions at the grand unification scale. In this
paper we shall investigate the implications of the non-universal
gaugino masses, as they arise in $SU(5)$ and $SO(10)$ grand unified
theories, on the neutralino and chargino mass 
spectrum~\cite{hep-ph/0211071, HLP, Profumo}.

The plan of this paper as follows. In section~\ref{sparticle}
we consider the renormalization group evolution of the sfermion masses
of the first- and second-generation in the case where the soft masses
are universal at the high scale, and then 
consider the evolution of these masses 
in $SU(5)$ and $SO(10)$ grand unified theories where these
masses are non-universal. We show how 
the determination of the sfermion masses can be used to determine
the soft parameters of the sfermion sector. The determination of these
parameters can then be used to distinguish between different
underlying supersymmetric grand unified theories. We then consider
interrelationships between the squark and slepton masses
in supersymmetric grand unified theories, and shown how
these can be used to distinguish between different breaking
patterns of a grand unified gauge group to the SM gauge group.
In section~\ref{nonuniv} we consider the non-universality
that can arise in the gaugino sector of a supersymmetric
grand unified theory, and its implications for the 
interrelationships between squark and slepton masses.
In section~\ref{neutchar} we consider the 
effect of gaugino non-universality on the neutralino and
chargino mass spectrum.  Finally, in section~\ref{thirdgen}
we consider the interrelationships between sfermion
masses of the third generation with universal boundary conditions,
as well as in grand unified theories
with non-universal soft masses. We conclude 
with a summary in section~\ref{summary}. In Appendix~\ref{rgeall} and
Appendix~\ref{SOLRGE}, we summarize the renormalization group
equations and their solutions for the general case which we use in
our paper.

\section{Sparticle spectroscopy in grand unified theories}
\label{sparticle}
In this Section we shall consider the sparticle spectrum 
in grand unified theories. More specifically, we shall consider 
the case when the underlying grand unified gauge group is 
either $SU(5)$ or $SO(10).$ 
With a given gauge group, the sparticle masses are obtained by 
the renormalization group evolution
of the soft supersymmetry breaking mass parameters from the 
GUT scale to the weak scale, with the boundary conditions on these
parameters at the GUT scale determined by the breaking pattern  of 
the grand unified gauge group to the Standard Model gauge group.
The underlying GUT group leaves its imprint on the sparticle spectrum 
through these boundary conditions. 
\subsection{Sfermion masses for first- and second-generation}
The renormalization group equations for the soft supersymmetry breaking 
squark and slepton mass parameters, the Higgs mass parameters, the trilinear
couplings, and the Yukawa couplings  are well known and are reproduced in the
Appendix~\ref{rgeall}.  
The renormalization group equations for the soft masses of 
squarks and sleptons of the first and second 
family~(the light generations) are especially simple, and are obtained  from
the renormalization group equations  (\ref{rg1}) -
(\ref{rg5}) by neglecting the Yukawa couplings, and by 
appropriately replacing other parameters. The RGEs for the light generations
can be solved analytically. Fortunately, we will not need the renormalization 
group equations for $m^2_{H_u}, m^2_{H_d}, A_t, A_b, A_\tau$ or the 
$\mu$ parameter for obtaining masses of the light generations. 
Since we can neglect the Yukawa couplings,
the light scalar particle spectrum consists of seven distinct groups of 
doubly degenerate scalar states 
$(\sul,\scl)$; $(\sdl,\ssl)$; $(\sur,\scr)$; $(\sdr,\ssr)$;
$(\sel,\smul)$; $(\ser,\smur)$; $(\seln,\smun)$. 
The seven third-generation sparticles will be split from the other 
two generations by the effects of Yukawa couplings.
The physical masses of the sfermions of the light
generations, which we shall denote by $M_{\tilde q}, M_{\tilde l}$
are easy to obtain.  These can be written as
\bea
\msuL &=& \msQO(t_G) +  
C_3(\tuL) + C_2(\tuL) + {1\over 36} C_1(\tuL)
+ ({1\over 2} - {2\over 3} \st) M_Z^2 \cos (2 \beta) - {1\over 5} K,
\label{sol1} \\
\msdL &=& \msQO(t_G) +  
C_3(\tdl) + C_2(\tdl)  + {1\over 36}  C_1(\tdl)
+ (-{1\over 2} +{1\over 3} \st) M_Z^2 \cos (2 \beta)- {1\over 5} K,
\label{sol2} \\
\msuR &=& \msuRO(t_G) + 
C_3(\tuR)                  + {4\over 9}  C_1(\tuR)
+ {2\over 3} \st M_Z^2 \cos (2 \beta) + {4\over 5} K,
\label{sol3} \\
\msdR & =& \msdRO(t_G) +  
C_3(\tdR)                 + {1\over 9} C_1(\tdR)
- {1\over 3} \st M_Z^2 \cos (2 \beta)- {2\over 5} K,
\label{sol4} \\
\mseL & =& \mslLO(t_G)                + C_2(\teL) + {1\over 4}   C_1(\teL)
+ (-{1\over 2} + \st)  M_Z^2 \cos (2 \beta)  + {3\over 5} K,
\label{sol5} \\
\msnL & = & \mslLO(t_G)               +  C_2(\tnL) + {1\over 4}  C_1(\tnL)
+ {1\over 2} M_Z^2 \cos (2 \beta) + {3\over 5} K,
\label{sol6}\\
\mseR & =& \mseRO(t_G)                            +   
C_1(\teR)
-\st M_Z^2 \cos (2 \beta) - {6\over 5} K, \label{sol7}
\eea
where $\tan\beta = v_u/v_d, v_u $ and $v_d$ being the vacuum 
expectation values of the two Higgs doublets of the minimal
supersymmetric standard model, and  where $C_1,\, C_2$ and $C_3$ are given by
\bea
C_i(t)= {a_i\over 2 \pi^2} \int_t^{t_G} dt~ g_i(t)^2~ M_i(t)^2, \, i=1,2,3,
\label{cidef} \\
a_1={3\over 5}, ~~~ a_2={3\over 4}, ~~~ a_3={4\over 3},
\eea
and
\bea
K = \frac{1}{16\pi^2}\int^{t_G}_t g_1^2(t)~S(t)~dt
  = \frac{1}{2b_1} \left[ S(t)- S(t_G) \right],
\label{Keqn}
\eea
is the contribution of the non-universality parameter $S$ to the 
sfermion masses. The non-universality parameter  $S$ is given by 
Eqs.~(\ref{sequation1}) --  (\ref{sequation2}) of 
Appendix~\ref{rgeall},  and $b_1=-33/5$. 
More explicitly
\bea
C_1(t) & = & {2 \over 11}{M_{1}^2(t_G)\over \alpha_G^2 }              
[\alpha_1^2(t_{G}) - \alpha_1^2(t)] \equiv  M_1^2(t_G) \overline{c}_1(t),\\
\label{Ic1eqn}\\
C_2(t) & = & {3 \over 2}{M_{2}^2(t_G)\over \alpha_G^2}
[\alpha_2^2(t_{G}) - \alpha_2^2(t)] \equiv  M_2^2(t_G) \overline{c}_2(t), 
\label{Ic2eqn} \\
C_3(t) & = & {8\over 9}   {M_{3}^2(t_G)\over \alpha_G^2} 
[\alpha_3^2(t) - \alpha_3^2(t_G)] \equiv M_3^2(t_G)  \overline{c}_3(t).
\label{Ic3eqn}
\eea
The sfermion masses (\ref{sol1}) - (\ref{sol7}) have contributions 
coming from different sources. First, there is the contribution coming 
from the mass at the large~(GUT) scale denoted by 
$m_{\tilde q_{L, R}}, m_{\tilde l_{L, R}}.$ 
Second, there is a contribution from the renormalization group~(RG)
running of scalar masses down to the  experimental scale.
Third, the contribution coming from $D^2$ term in the scalar potential,
which is proportional to $M_Z^2.$ Finally, there is the
contribution from the non-universality of the sfermion masses, which is
proportional to $K$. The contributions coming from 
corresponding quark and lepton masses is completely negligible for 
all the sfermions of the light generations, as is the contribution
from the sfermion mixings.  We note that in arriving at (\ref{sol1}) --
(\ref{sol7}), we have integrated the RGEs 
to the physical sparticle mass to obtain the $C_i's.$  
Since the sfermion masses involve the values of soft masses at the 
initial scale~(GUT scale), their values at the experimental scale
will be determined by their values at the GUT scale. 
The simplest case is that of universal soft masses at the 
initial GUT scale:
\bea
\msQO(t_G)  & = &  \msuRO(t_G) = \mseRO(t_G) 
= \mslLO(t_G) = \msdRO(t_G) = m_0^2,  \label{sfcond0}\\
M_1^2(t_G) & = & M_2^2(t_G)= M_3^2(t_G)= M_{1/2}^2 \label{gaugino0}, 
\eea
for which $S = K = 0.$ In this case all the sfermion masses can be expressed 
in terms of three parameters only. These are  $m_0^2,  M_{1/2}^2$
and $\cos(2\beta).$ These three parameters can be determined from a
measurement of three of the sfermion masses, say, for example, 
$\tilde u_L, \tilde d_L,$ and $\tilde e_R$~\cite{FHKN}. 
We have from Eqs.~(\ref{sol1}) - (\ref{sol7}) 
\bea
\msq & = & {1\over X_U} \left[ 3(\cdL+\cuL)\mseR-3\ceR(\msdL+\msuL) \right. 
\nonumber \\
 && \left.
+2 ( -2\cdL \mseR- \cuL(3\msdL+\mseR) + 3\cdL \msuL 
+\ceR(2 \msdL+\msuL) )\stwoW \right], \label{u1} \\
M_{1/2}^2 & = & {1 \over X_U}
\left[ 3(\msdL-2\mseR+\msuL)+2(\msdL+3\mseR-4\msuL)\stwoW \right], \label{u2} \\
\ctwoB & = &
-{{1} \over {X_U \MZsq}} 
\left[ 6 \cuL(\msdL-\mseR)+\ceR(\msuL-\msdL)+\cdL(\mseR-\msuL)\right], 
\label{u3}
\eea
where
\bea
X_U & = & 3(\cdL-2\ceR+\cuL)+2(\cdL+3\ceR-4\cuL) \stwoW,\\
\cdL & = &  \cthree(\tdl) + \ctwo(\tdl)  + {1 \over 36} \cone(\tdl), \\
\cuL & = & \cthree(\tuL) + \ctwo(\tuL) + { 1 \over 36} \cone(\tuL), \\
\ceR & = & \cone(\teR).
\eea
Measurement of the remaining four doubly degenerate sparticle masses
overdetermines the unknown parameters. This allows for a check of the
consistency of underlying assumption of universal soft 
mass parameters. If the seven doubly degenerate sparticle masses
cannot be fit with three parameters, then new physics beyond
that of universal soft masses at the large scale must be invoked.

\begin{figure}[!ht]
\psfrag{muL}[r,t][r,t]{$M_{\tilde{u}_L}$}
\psfrag{meR}[l,t][b,t]{$M_{\tilde{e}_R}$}
\psfrag{m0}[r,t][l,t]{$m_{0}$}
\includegraphics[width=4.7cm]{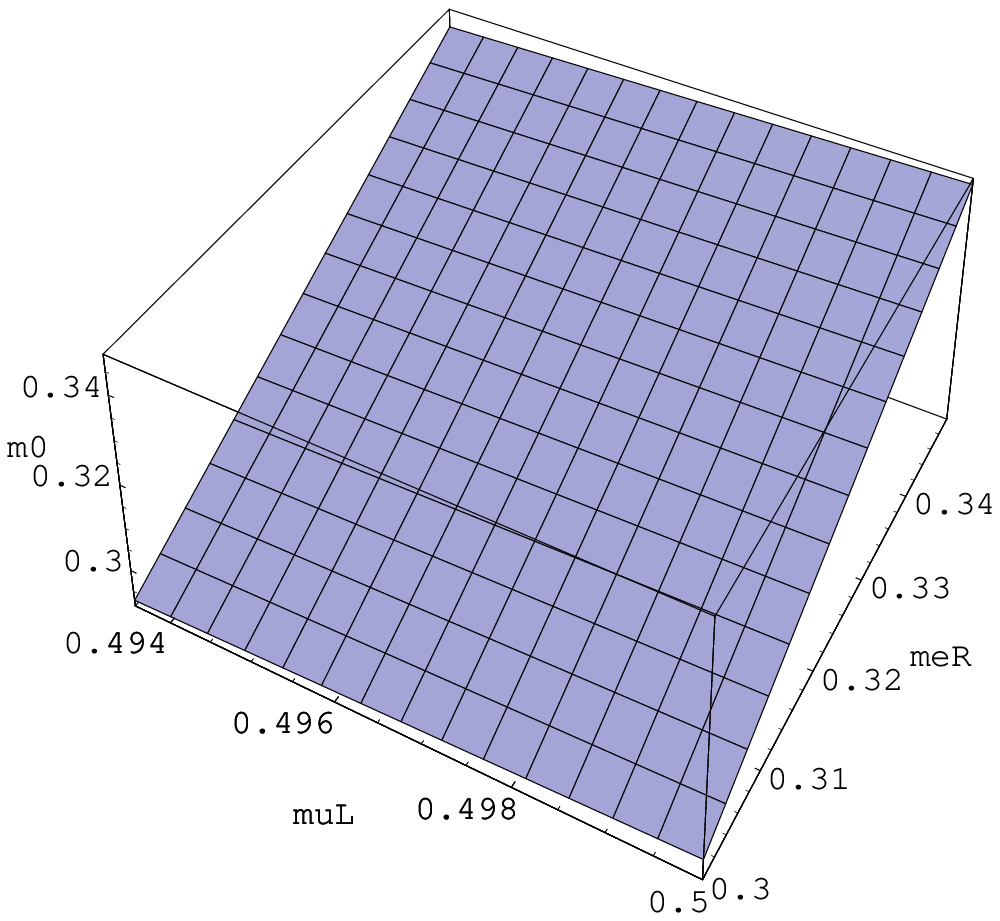}\hskip 1.5cm
\psfrag{muL}[r,t][r,t]{$M_{\tilde{u}_L}$}
\psfrag{meR}[l,b][r,t]{$M_{\tilde{e}_R}$}
\psfrag{M0}[r,t][l,t]{$M_{1/2}\ \ $}
\includegraphics[width=4.7cm]{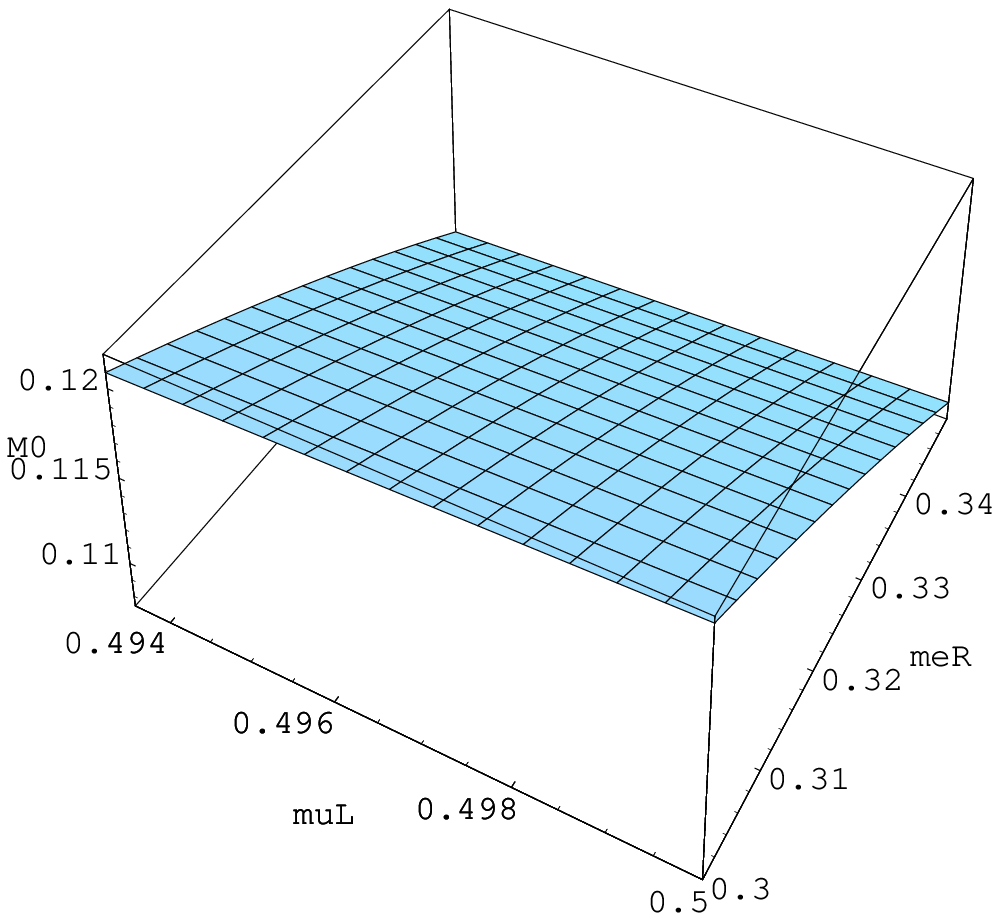} \hskip 1.5cm
\psfrag{muL}[r,t][r,t]{$M_{\tilde{u}_L}$}
\psfrag{meR}[l,t][l,t]{$M_{\tilde{e}_R}$}
\psfrag{c2B}[r,t][l,t]{$\ctwoB$}
\includegraphics[width=4.7cm]{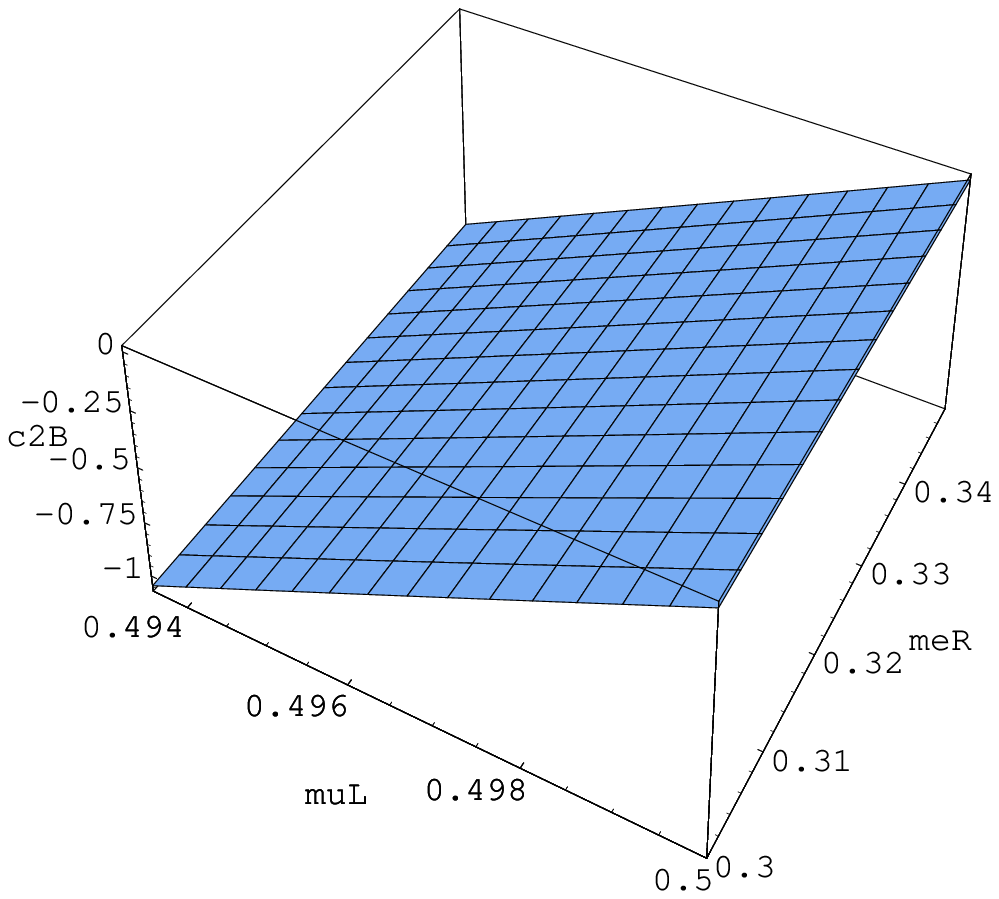}
\caption{The parameters of the supersymmetric standard model with
universal boundary conditions as a function of the sparticle
masses. The first frame  shows  the universal scalar mass parameter
$m_0$ as a function of $M_{\tilde{u}_L}$ and $M_{\tilde{e}_R}$,
with  $\mdL=0.5$. All masses are in TeV. The second frame 
shows  $M_{1/2}$ as a function of sparticle masses.  All parameters
are as in the first frame.  The third frame shows
$\ctwoB$ as a function of sparticle masses, with parameters as
in the first plot.}
\label{universalfig}
\end{figure}

In Fig.~\ref{universalfig}
we plot the parameters $m_0, M_{1/2}$ and $\cos(2\beta)$ 
for the case of universal boundary conditions (\ref{sfcond0}) and
(\ref{gaugino0})
as a function of sparticle masses.
In this Fig. we have taken $M_{\tilde{d}_L}$ fixed at 0.5 TeV, and
varied $M_{\tilde{u}_L}$ in a range dictated by the requirement that
$\ctwoB$ lies in the range of $-1 \leq
\ctwoB \leq 0$ (we take $\tan\beta \ge 1)$).
The mass of the selectron $M_{\tilde{e}_R}$ is varied
over a reasonable range for purposes of illustration. 
We have chosen what we consider to be a typical set of values.
It may be observed that the effects due to $\ctwoB$ leads to
a very small splitting between the $u_L$ and $d_L$ squarks
and there is no known mechanism to lift the near degeneracy any further.
The degeneracy would be exact here, and in the case of non-universal
boundary conditions which are to be discussed later, but for
electroweak symmetry breaking effects.  Clearly larger values of
$M_{\tilde{e}_R}$ lead to larger values of $m_0$, while they
lead to smaller values of $M_{1/2}$ for the other two squark masses
held fixed, which is dictated by the renormalization group flow.
Once these three parameters are determined from the measurements
of these three sparticle masses, they may be inserted back into
the expressions for the other sparticle masses and  compared
with the experimental values of these other  masses.  The predictions
of universal boundary conditions may be readily studied in this manner.
We will see, however, in the next section that the discussion becomes
significantly more complicated for the case of non-universal boundary
conditions which occur in  grand unified theories.

\subsubsection{ The $SU(5)$ supersymmetric grand unified theory}
Supersymmetric models with universal squark and slepton mass parameters
at the large scale are economic in terms of new parameters, but 
could to some extent be unrealistic.
One of the realistic  models which goes beyond the assumption
of universality of soft scalar masses 
is the  $SU(5)$ supersymmetric GUT.  
Since $Q_L, u_R, e_R$ all lie in a $10$-dimensional representation, and 
$L_L, d_R$ lie in a $\bar 5$  representation of  $SU(5)$, 
the boundary conditions for the soft masses for squarks and sleptons 
can be written as 
\bea
\msQO(t_G) & = & \msuRO(t_G) = \mseRO(t_G) = m_{\bf 10}^2, \label{5cond1}\\
\mslLO(t_G) &=& \msdRO(t_G) = m_{\bf \overline{5}}^2.             
\label{5cond2}
\eea
Here  $m_{\bf 10}^2$ and $m_{\bf \overline{5}}^2$
are the common squared soft scalar masses corresponding to
the $\bf 10$- and $ \overline{\bf 5} $-dimensional representations, 
respectively of $SU(5)$, at the unification scale.
For the $SU(5)$ supersymmetric grand unified theory, the universal gaugino mass condition (\ref{gaugino0}) continues to hold.
We note that the two Higgs doublets
of the minimal supersymmetric standard model  belong to two different
representations of $SU(5)$, and as such  their soft parameters 
are unrelated to each other. Since, we don't need the Higgs mass
parameters in our calculations, we do not write these explicitly.

The boundary conditions
(\ref{5cond1}) and (\ref{5cond2}) are valid for each generation. 
Since there is no boundary condition relating masses of particles 
in different generations, we have  four input parameters, namely,
$\mfivesq, \mtensq, M_{1/2},$ and $\cos(2\beta)$, in terms of which the squark
and slepton masses of each light generation can be computed. These input
parameters, together with the non-universality parameter $K$,
can be determined by a measurement of five sfermion masses. 
Taking  these to be $\tilde u_L, \tilde d_L,  \tilde d_R,   \tilde u_R$
and $\tilde e_R,$ we have  from Eqs.~(\ref{sol1}) - (\ref{sol7})
\bea
\mfivesq & = & {1 \over 5 X_5} \left[ -\cuR(\msdL+5 \msdR - 2 \mseR + \msuL) +
                          \ceR(\msdL-5 \msdR +\msuL - 2 \msuR) \right .
			  \nonumber \\
& & \left . - 5\cdR(\msdL - \mseR + \msuL - \msuR) 
    +  (\cdL+\cuL) (5 \msdR - \mseR + \msuR)\right], \\
\mtensq & = & {1 \over 5 X_5} \left[\cuR(-3\msdL + \mseR - 3 \msuL)
                - \ceR (2(\msdL + \msuL) + \msuR)
		+ (\cdL+\cuL)(2\mseR+3\msuR)\right], \\
M_{1/2}^2 & = & {1 \over X_5} \left[ \msdL - \mseR + \msuL - \msuR \right], \\
\ctwoB & = & {{1} \over {X_5 \MZsq (\stwoW-1)}}
             \left[ \ceR(\msuL-\msdL)+\cuR(\msuL-\msdL)
             + \cdL(\mseR-2\msuL + \msuR)\right],
\eea
where
\bea
X_5 & = &  \cdL-\ceR+\cuL-\cuR ,\\
\cuR & = & \cthree(\tuR)+ { 4 \over 9}  \cone(\tuR),\\
\cdR & = & \cthree(\tdR) + {1 \over 9} \cone(\tdR).
\eea
Since the two Higgs doublet
superfields $H_d$ and $H_u$ of the MSSM lie in different representations of
$SU(5),$ $m^2_{H_d}(t_G)$ and   $m^2_{H_u}(t_G)$ are not  equal.
Thus, the parameter 
\bea
S(t_g) & = & m_{H_u}^2(t_g) - m_{H_d}^2(t_g),
\eea
is nonzero in this case. This parameter, or equivalently 
the parameter $K$, can be expressed in terms of the sfermion masses as
\bea
K & = & {{1}\over {6 X_5 (\stwoW-1)}} \left[ 3\cuR(\msdL - 2 \mseR + \msuL)
+ 3 (\cdL + \cuL)(\mseR - \msuR) \right . \nonumber \\
&& \left . + 2\left(\cuR(\msdL+3 \mseR - 4 \msuL)
 - \cdL(4 \mseR - 5 \msuL + \msuR) +
\cuL(-5 \msdL + \mseR + 4 \msuR)\right) \stwoW \right . \nonumber \\
&& \left . + \ceR \left( -3(\msdL+\msuL-2 \msuR) 
- 2(-4\msdL+\msuL+3\msuR)\stwoW \right)
\right]. 
\eea

\begin{figure}[!ht]
\psfrag{muL}[r,t][r,t]{$M_{\tilde{u}_L}$}
\psfrag{meR}[l,t][b,t]{$M_{\tilde{e}_R}\ \ $}
\psfrag{m5}[r,t][l,t]{$m_{\bf \overline{5}}\ $}
\includegraphics[width=5cm]{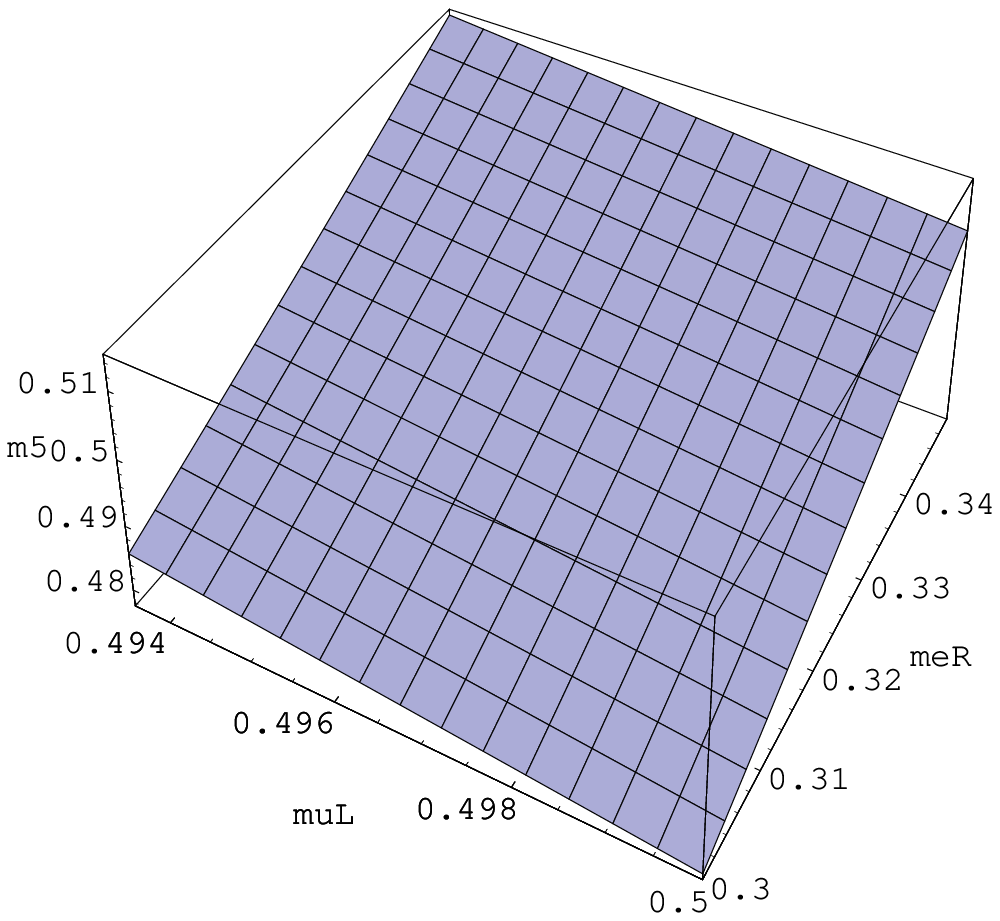} \hskip 1cm
\psfrag{muL}[r,t][r,t]{$M_{\tilde{u}_L}$}
\psfrag{meR}[l,t][b,t]{$M_{\tilde{e}_R}\ \ $}
\psfrag{m10}[r,t][l,t]{$m_{\bf 10}\ $}
\includegraphics[width=5cm]{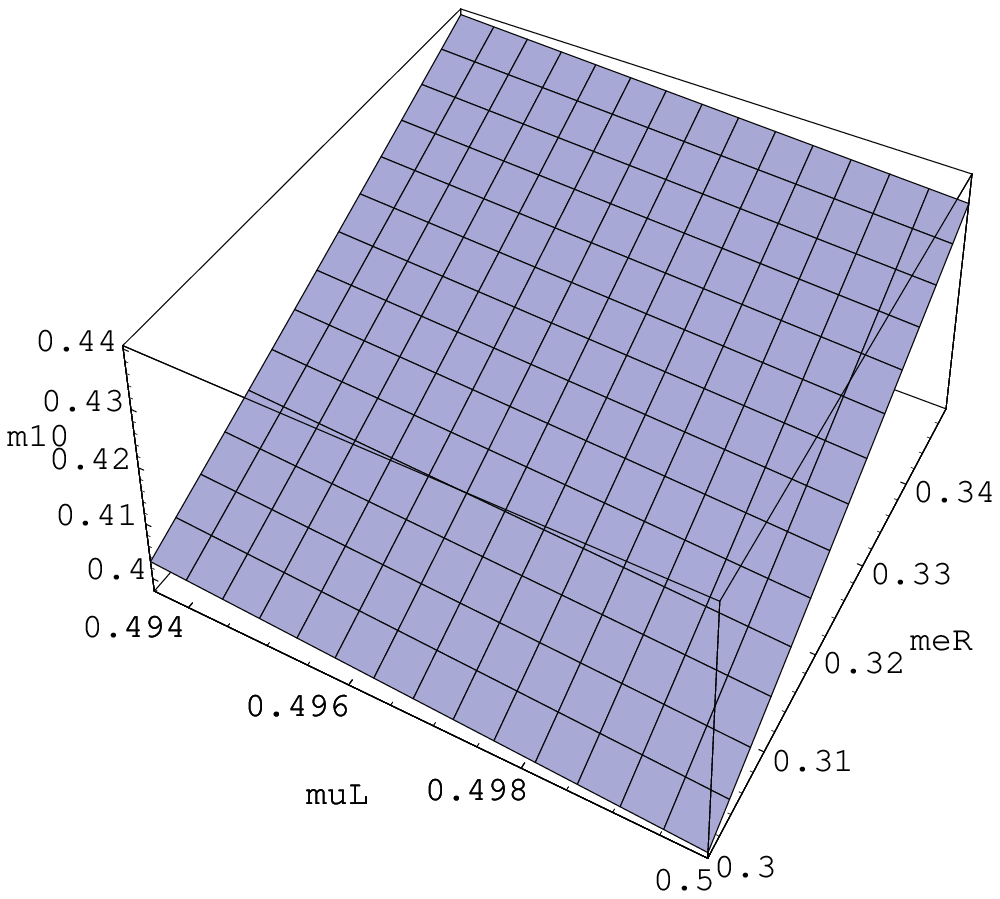}  \hskip 1cm 
\psfrag{muL}[r,t][r,t]{$M_{\tilde{u}_L}$}
\psfrag{meR}[l,t][l,t]{$M_{\tilde{e}_R}\ \ \ $}
\psfrag{M0}[r,t][l,t]{$M_{1/2}$}
\includegraphics[width=5cm]{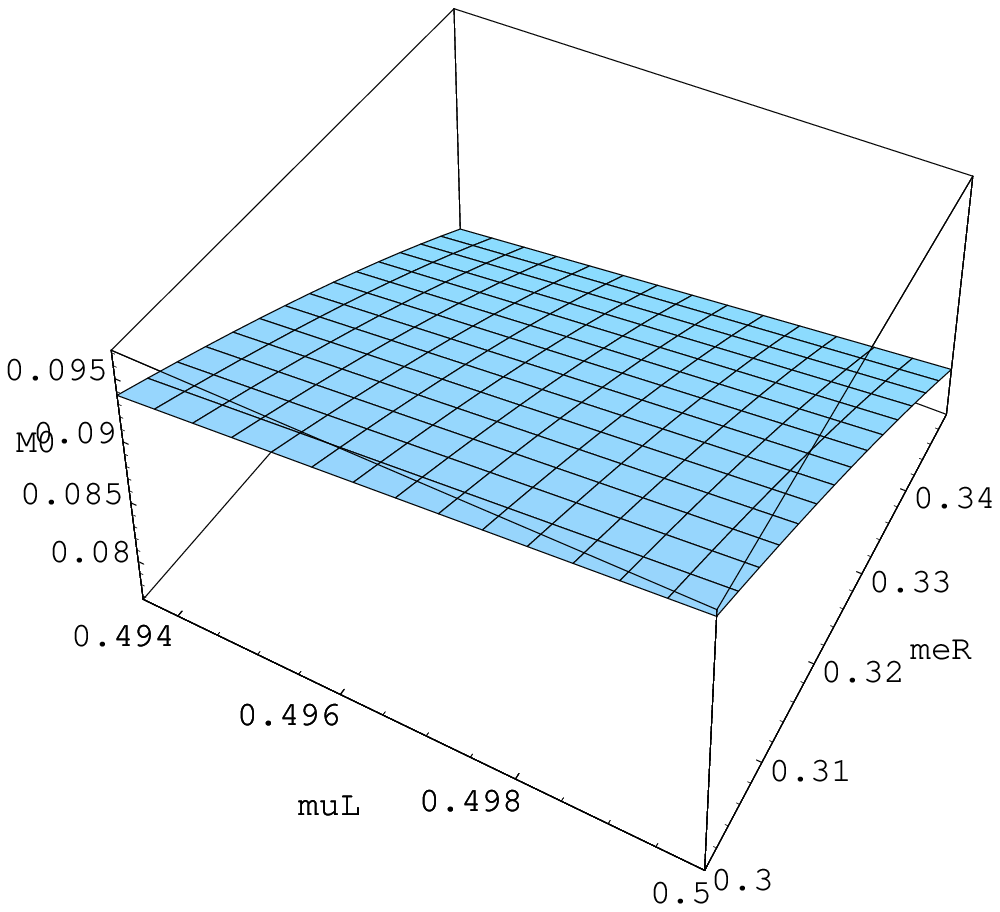} \\
\psfrag{muL}[r,t][r,t]{$M_{\tilde{u}_L}$}
\psfrag{meR}[l,t][l,t]{$M_{\tilde{e}_R}\ \ \ $}
\includegraphics[width=5cm]{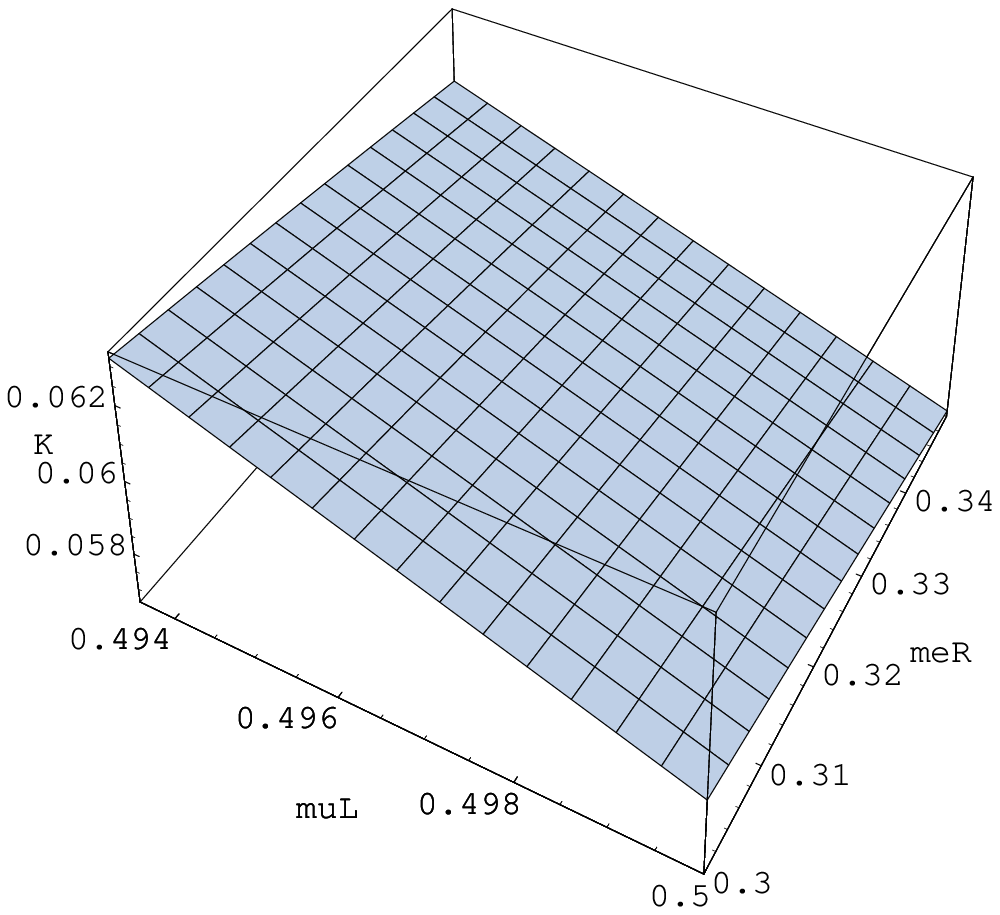} \hskip 1.5cm
\psfrag{muL}[r,t][r,t]{$M_{\tilde{u}_L}$}
\psfrag{meR}[l,t][l,t]{$M_{\tilde{e}_R}\ \ \ $}
\psfrag{c2B}[r,t][l,t]{$\ctwoB$}
\includegraphics[width=5cm]{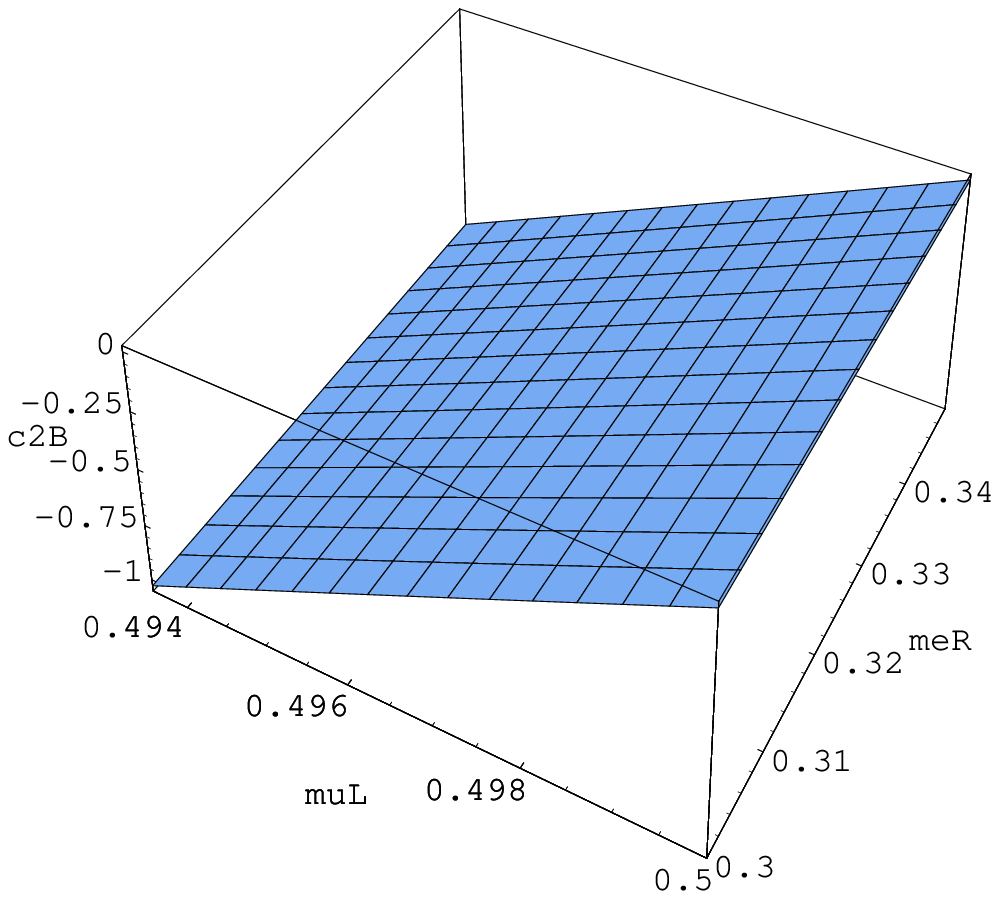} 
\caption{The parameters of $SU(5)$ supersymmetric grand unified theory
as a function of sparticle masses. The first frame shows 
the plot of $m_{\bf \overline{5}}$ as a function of  $\muL$ and $\meR$,
with $\mdL=0.5,\mdR=\muR=0.55$, with all masses in TeV. 
The second frame  shows the plot of $m_{\bf 10}$, whereas
the third, fourth, and fifth frames  show
the variations of $M_{1/2}$, $K$, and  $\ctwoB$ as a function of 
sparticle masses. All parameters are as in the first frame.}
\label{su5fig}
\end{figure}
In Fig.~\ref{su5fig}
we plot the parameters $m_{\overline 5}, m_{10}, M_{1/2}^2, 
K$ and $\cos(2\beta)$. 
We have taken $M_{\tilde{d}_L}$ fixed at 0.5 TeV,
and chosen $M_{\tilde{d}_R}=M_{\tilde{u}_R}=0.55$ TeV
and varied $M_{\overline{u}_L}$ in a range dictated by the requirement that
$\ctwoB$ lies in the range of $-1 \leq
\ctwoB \leq 0$ as before.                     
The mass of the selectron $M_{\tilde{e}_R}$ is varied
over a reasonable range for purposes of illustration. 
Note that in the case of universal boundary conditions we would
expect to see the right handed squarks to be somewhat lighter than
the left handed squarks, and also that in that case exact degeneracy
would not be possible.  Here it would be possible to have
exact degeneracy of the states due to the interplay between the
non-universality parameter contributions balancing the electroweak
contribution.  Such a possibility, although highly simplified
makes our discussion simple and is useful for purposes of illustration.
We have chosen what we consider to be a typical set of values.
It may be observed that the effects due to $\ctwoB$ leads to
a very small splitting between the $u_L$ and $d_L$ squarks
and there is no mechanism to lift the near degeneracy any further.
It may be from our expressions 
that a spectrum of the sort depicted here
requires a large splitting between $m_{\overline{5}}$ and $m_{10}$,
whereas $M_{1/2}$ remains relatively less sensitive.

\subsubsection{ The $SO(10)$ supersymmetric grand unified theory}
As discussed above, the simplest grand unified theory which results
in non-universal masses for the sfermions is $SU(5)$ GUT. 
The rank of $SU(5)$ is the same as that of SM gauge group.
On the other hand SM can be embedded into a larger gauge group 
like $SO(10)$. However, since the rank of $SO(10)$ is higher than
that of SM gauge group, the breaking of $SO(10)$  to the SM gauge group
involves the reduction of rank by one unit.  There is an additional 
$U(1)$ factor  beyond that of the Standard Model, which must be broken.
Thus, in the case of $SO(10)$ unfied theory there will be $D$-term 
contributions to the soft supersymmetry breaking scalar masses.
These  $D$-term contributions have important phenomenological 
consequences at low energies as these allow one to reach
regions of parameter space  which are not otherwise  accessible
with universal boundary conditions. These contributions can help
distinguish between  different scenarios for breaking
of grand unified symmetry at high energies.

The   $D$-term contributions to the soft scalar masses will depend
on the manner in which $SO(10)$ gauge group is broken to
the SM gauge group. When $SO(10)$ breaks via its maximal
subgroup $SU(5)\times U(1)_Z$, with
$SU(5) \supset SU(3)_C\times SU(2)_L \times U(1)_X$, there are
two possibilities for the hypercharge generator of the
SM gauge group.
In the ``conventional'' embedding via $SU(5)$,
the hypercharge generator $Y$ of the SM is identified with the generator
$X$ of $U(1)_X$. On the other hand, in the ``flipped'' embedding
the hypercharge generator is identified with a linear combination of
the generators $X$ and $Z$.

Apart from the ``natural'' subgroup
$SU(5)\times U(1)$, the group $SO(10)$ also has ``natural''
subgroup $SO(6)\times SO(4).$
Since $SO(6)$ is isomorphic to $SU(4)$, and $SO(4)$  is isomorphic
to $SU(2) \times SU(2)$, $SO(10)$ contains~\cite{Pati:1974yy} the group
$SU(4)\times SU(2) \times SU(2)$. We shall focuss on the
signatures of the  $SO(10)$ breaking to the SM 
via its two natural subgroups, and try to find distinguishing features 
of the sparticle spectrum in the two cases.

As in the case of $SU(5)$ GUT, the solutions of the RG equations for the 
soft scalar masses involve the values of these masses at the initial 
GUT scale.  These initial values will be determined by the 
pattern of the breaking of the grand unified  group to the SM gauge group. 
For the case of breaking of $SO(10)$  to the Standard Model gauge group
via its maximal subgroup $SU(5)\times U(1)$, 
these initial values are given by~\cite{Kawamura1,Kawamura2,AP2,AP1}
\bea
\msQO(t_G) &=& \msuRO(t_G) = \mseRO(t_G) = m_{\bf 16}^2+
g_{10}^2D, \label{cond1}\\
\mslLO(t_G) &=& \msdRO(t_G) = m_{\bf 16}^2-3g_{10}^2D,  \label{cond2}\\
\mhUO(t_G) &=& m_{\bf 10}^2-2g_{10}^2D, \label{cond3}\\
\mhDO(t_G) &=& m_{\bf 10}^2+2g_{10}^2D.\label{cond4}
\bigskip
\eea
at the $SO(10)$ breaking scale $M_G$, where the normalization and sign
of $D$ is arbitrary.
Here $m_{\bf 16}$ and  $m_{\bf 10}$ are the common soft scalar masses,
corresponding to the $\bf{16}-$ and $\bf{10}-$ dimensional representations,
respectively of $SO(10)$, at the unification scale, and $g_{10}$ is the
$SO(10)$ gauge coupling.
We note here that in the breaking of
$SO(10)$ the rank is reduced by one, and hence the $D$-term contribution
to the soft masses is expressed by a single parameter $D$.
We also note that   $g_{10}$ and $D$ enter the boundary conditions
(\ref{cond1}) -- (\ref{cond4}) in the combination $g_{10}^2 D$, and,
therefore, constitute only one parameter.
Thus, in the case of $SO(10)$ grand unified theory, there are 
four input parameters, the same as in the case of $SU(5)$ grand 
unified theory, which determine the light sfermion sector.
We can take these to be
$m_{16}^2$, $g_{10}^2D$,  $M_{1/2}$, and $\cos(2\beta)$.
In addition there is the non-universality parameter $K$.
Using the solutions (\ref{sol1}) --  (\ref{sol7}) of the RG equations,
these parameters can be determined 
in terms of squark and slepton masses as
\bigskip
\bea
 \msixteensq & = & {1\over 4 X_5} \left[ -\cuR(2\msdL+\msdR-\mseR+2\msuL) +
                \cdR(-\msdL+\msdR-\msuL+\msuR) - \right. \nonumber \\
& & \left. \ceR(\msdL+\msdR+\msuL+\msuR)+(\cdL+\cuL)(\msdR+\mseR+2\msuR) \right]
 \\
g_{10}^2 D & = & {1\over 20 X_5} \left[-\cuR(2\msdL-5\msdR+\mseR+2\msuL)-
		(\cdL+\cuL)(5\msdR-3\mseR-2\msuR)  \right. \nonumber \\
& & \left. 
+ 5\cdR(\msdL-\mseR+\msuL-\msuR)+\ceR(-3 \msdL + 5 \msdR 
- 3 \msuL +\msuR)\right].
\label{so10relations}
\eea
All other quatities are same as in the $SU(5)$ case.

\begin{figure}[htb]
\psfrag{muL}[r,t][r,t]{$M_{\tilde{u}_L}$}
\psfrag{meR}[l,t][l,t]{$M_{\tilde{e}_R}\ \ \ $}
\psfrag{m16}[r,t][l,t]{$m_{\bf 16}$}
\includegraphics[width=6cm]{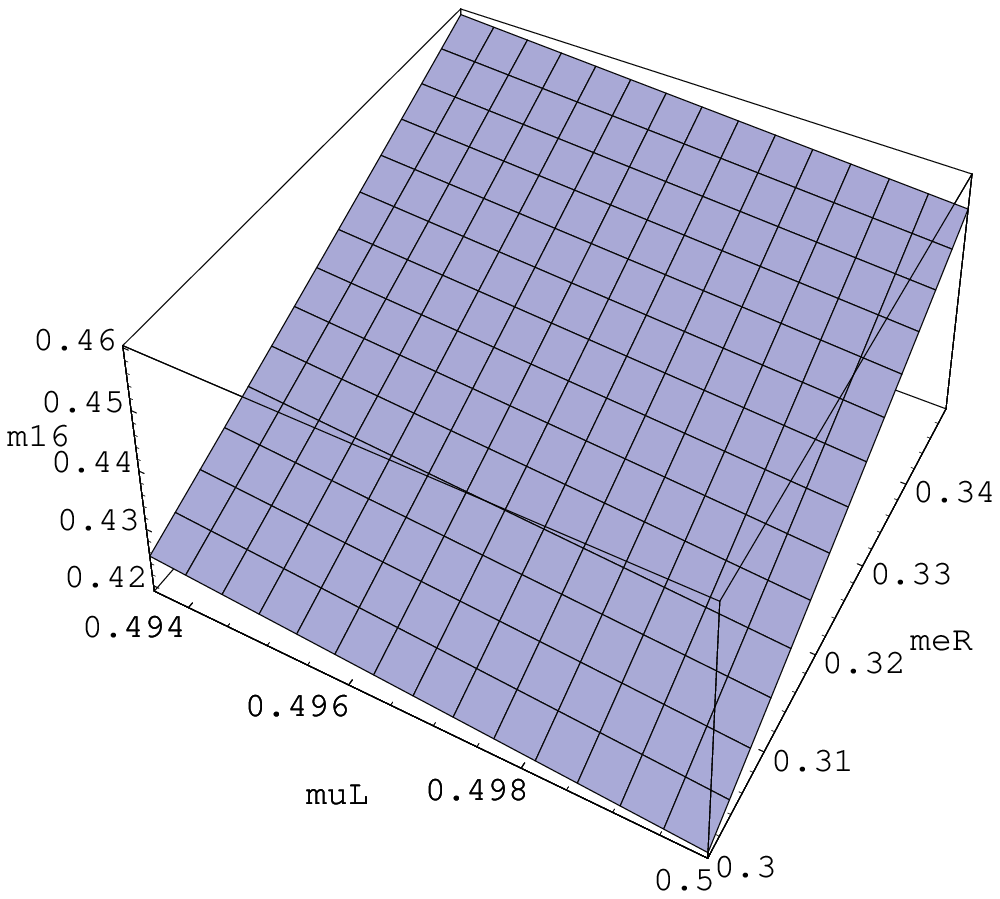} \hskip 1.5cm
\psfrag{muL}[r,t][r,t]{$M_{\tilde{u}_L}$}
\psfrag{meR}[l,t][l,t]{$M_{\tilde{e}_R}\ \ \ $}
\psfrag{DD2}[r,t][l,t]{$g_{10}^2 D~~~$}
\includegraphics[width=6cm]{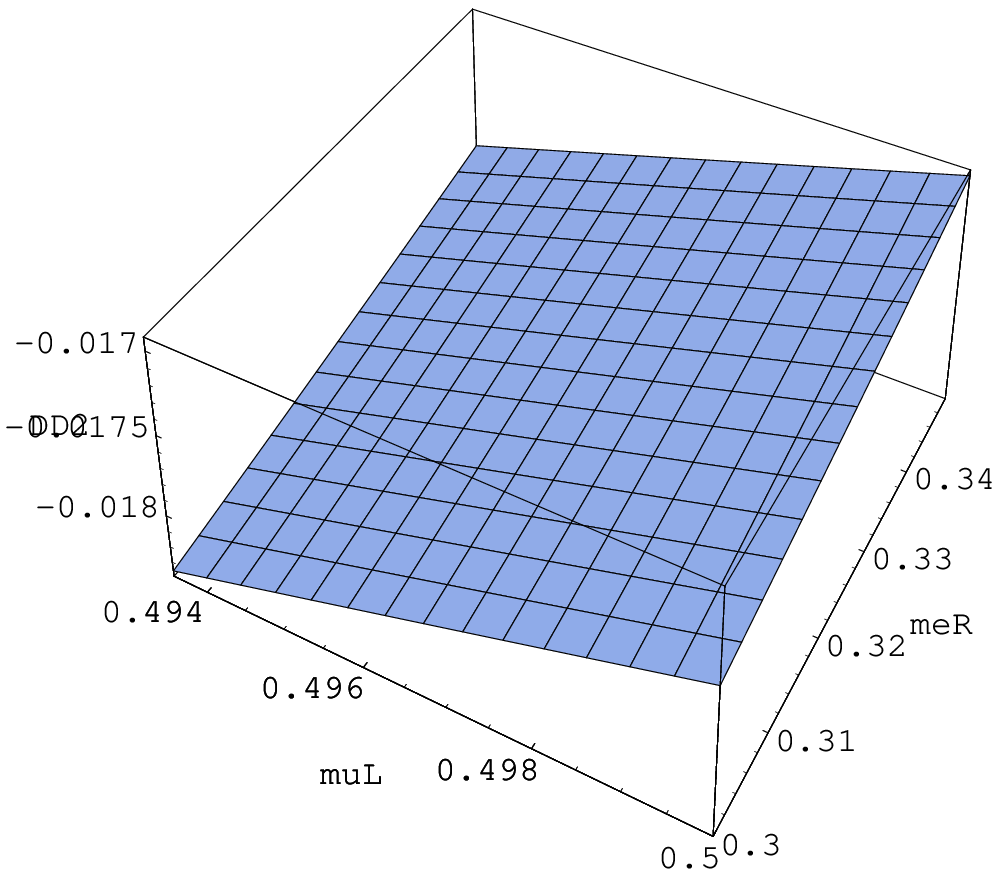}
\caption{The parameters of $SO(10)$ supersymmetric grand unified theory
as a function of sparticle masses. The first frame shows 
the plot of $m_{\bf \overline{16}}$ as a function of  $\muL$ and $\meR$,
with $\mdL=0.5,\mdR=\muR=0.55$, with all masses in TeV. 
The second frame shows the plot of $g_{10}^2 D$ as a function of 
sparticle masses. All parameters are as in the Fig. 2}
\label{so10fig}
\end{figure}
In Fig.~\ref{so10fig} we show the variation of
the $SO(10)$ parameters  $m_{\bf 16}$ and $g_{10}^2 D$ as a 
function of sfermion masses. The rest of the parameters $M_{1\over 2},  
\cos 2\beta$ and $K$ for SO(10) are same as as in the case of and $SU(5)$, 
and are not shown here.

It is important to note that once  $\msixteensq $ and $g_{10}^2 D$ 
are obtained, the boundary conditions for all the squark and slepton soft
masses are fixed through (\ref{cond1}) -- (\ref{cond2}). The masses of
of all the first and second generation squarks 
can then be obtained from the 
solutions (\ref{sol1}) - (\ref{sol7}) of the RG equations.

Another important aspect of (\ref{cond1}) -- (\ref{cond4}) is that the 
boundary condition for the 
difference of the Higgs mass parameters is given by
$S(t_G) = \mhUO(t_G) - \mhDO(t_G) = -4g_{10}^2D$. This quantity
can, thus, be  determined from the measurement of sfermion masses of the
light generations through the relation (\ref{so10relations}). 
This implies  that the value  of
the difference of the Higgs mass parameters at the electroweak
scale, which is
crucial for the  electroweak symmetry breaking, can be 
obtained from the difference of the RG equations (\ref{rg7}) and
(\ref{rg8}) with the boundary condition determined
by the observed masses of the sparticles through
the relation (\ref{so10relations}). This is different from what 
happens in $SU(5)$, where  the difference of the Higgs mass 
parameters at the electroweak
scale cannot be obtained in terms of the sparticle masses alone.

We now come to the case of $SO(10)$ breaking via its other maximal
subgroup $SO(10) \supset SU(4)_{PS} \times SU(2)_L \times SU(2)_R.$
As in the case of breaking via the $SU(5)\times U(1)$ subgroup, there
appear $D$-term contributions to the soft scalar masses when the rank
of the gauge group reduces from 5 to 4 at the intermediate Pati-Salam
symmetry breaking scale $M_{PS}$.
For this  case of breaking, the initial values of soft masses
are given by~\cite{AP2,AP1}
\begin{eqnarray}
\msQO(M_{PS}) &=& m_L^2 + g_4^2 D, \label{ps1}\\
\msuRO(M_{PS}) &=& m_R^2 - (g_4^2 - 2 g_{2R}^2) D,  \label{ps2}\\
\mseRO(M_{PS}) &=& m_R^2 + (3 g_4^2 - 2 g_{2R}^2) D,  \label{ps3}\\
\mslLO(M_{PS}) &=& m_L^2 - 3 g_4^2 D, \label{ps4} \\
\msdRO(M_{PS})  &=& m_R^2 - (g_4^2 + 2 g_{2R}^2) D,  \label{ps5}
\end{eqnarray}
at the Pati-Salam breaking scale. Here $D$ represents
the $D$-term contributions whose normalization is
arbitrary. We note that the boundary conditions (\ref{ps1}) -- (\ref{ps5})
do not depend on a particular
choice of the Higgs representation which breaks the Pati-Salam group,
but is fixed only by the symmetry breaking pattern.
Here, $m_L, m_R$ are the soft masses corresponding
to the $SU(2)_L, SU(2)_R$ gauge groups, 
and $g_4, g_{2R}$ are the 
$SU(4)_{PS}, SU(2)_R$ gauge couplings, repectively. Furthermore,
the initial soft sfermion masses are parametrized by four parameters, which we
take to be  $m_L^2,  m_R^2,  g_4^2 D,$ and $g_{2R}^2) D$, respectively. 
Also the gauge coupling $g_4^2$, $g_{2R}^2$ can be determined
from the low-energy gauge coupling $\alpha_i (m_Z)$ $(i=1,2,3)$ as
a function of $M_{PS}$ alone. 

The four parameters  $m_L^2,  m_R^2,  g_4^2 D,$ and $g_{2R}^2 D$ together
with  $M_{1/2}$, $K$ and $\cos(2\beta)$ constiutute a set of
seven parameters in terms of which the light sfermion mass
spectrum can be determined. However, the system of 
equations which determine these seven parameters
in terms of  seven sparticles masses 
is degenerate and hence no
solution exists. Thus, in this case we cannot determine the
underlying parameters in terms of the sfermion masses alone.

\subsection{Sum Rules}

The fact that the renormalization group equations for the
soft mass parameters for the first two generations can be
solved analytically allows us to determine the 
parameters of the underlying  supersymmetric theory.
This can also allow us to distinguish between 
different grand unified gauge groups.
In this subsection we show that in
grand unified theories, the squark and slepton masses
obey certian relations, which follow
from the renormalization group evolution and
the boundary conditions at the GUT scale.

For the case of non-universal  boundary conditions for the scalar masses
which obtain in both SU(5) and SO(10) models,
we eliminate  
$m_{\tilde q}, m_{\tilde l}$ from the solutions of the 
renormalization group equations of the soft scalar masses (\ref{sol1})
- (\ref{sol7}) to obtain the following two sum rules for the sfermion masses:
\bea
2 M_{\tilde{Q}}^2 - \msuR -\mseR &=& (C_3 + 2C_2 - {25\over 18}C_1),
\label{Isum1}\\
M_{\tilde{Q}}^2 + \msdR - \mseR - m_{\tilde{L}}^2
&=& (2C_3 - {10\over 9} C_1)\label{Isum2},
\eea
where we have used the notation
\bea
M_{\tilde{Q}}^2={1\over 2} (\msuL + \msdL),  \, \,  \, M_{\tilde{L}}^2
={1\over 2} (\mseL + \msnL). \nonumber
\eea
These sum rules are valid  for $SU(5)$ supersymmetric grand unified
theory. Although, the rank of $SO(10)$ gauge group is larger than
the SM (and $SU(5)$), these sum rules are also valid in a
supersymmetric $SO(10)$ grand unified theory when
$SO(10)$ breaks to the
SM gauge group via its maximal $SU(5) \times U(1)$ subgroup, 
irrespective of whether the embedding of the SM in $SU(5) \times U(1)$
is  conventional or flipped one.

On the other hand  in the case of $SO(10)$ breaking via its 
other maximal subgroup 
$SO(10) \supset SU(4)_{PS} \times SU(2)_L \times SU(2)_R$,
using the boundary conditions (\ref{ps1}) - (\ref{ps5}), we
obtain the sum rule
\bea
m_{\tilde{Q}}^2 + \msdR - \mseR - m_{\tilde{L}}^2
&=& (2C_3 - {10\over 9} C_1)\label{Isum3},
\eea
which is the only sum rule valid in this case. 
Thus, this sum rule serves as a crucial
distinguishing feature of $SO(10)$ breaking via the Pati-Salam subgroup.
If both the sum rules~(\ref{Isum1}) and ~(\ref{Isum2}) are seen to hold
experimentally, then in the context of grand unification,
the underlying gauge group is either $SU(5)$ or $SO(10)$,
with the breaking
of $SO(10)$ taking place via its $SU(5)$ subgroup. On the 
other hand, if only
the sum rule~(\ref{Isum3}) is seen to hold experimentally, then the breaking
of $SO(10)$ must take place  via the Pati-Salam subgroup.

We now turn to the  
the right hand side of the sum rules (\ref{Isum1}) and (\ref{Isum2})
which involve the functions $C_i$.
These functions can be written in terms of quantities whose values can be
inferred from experiment. In terms of the gluino mass $M_{\tilde g}
= M_3(t_{\tilde g})$, we can write $C_3(t)$  from (\ref{Ic3eqn}) as
\be
C_3(t) =  {8\over 9}   {M_{\tilde g}^2\over \alpha_3^2(t_{\tilde g})} 
[\alpha_3^2(t) - \alpha_3^2(t_G)], \label{Nc3eqn}\\
\ee                                                                           where we have used the fact that gaugino masses run as 
\be
{M_i(t)\over \alpha_i(t)} = {M_i(t_G)\over \alpha_i(t_G)}.
\label{Igauginoeq}    
\ee
We then have
\be
{M_1(t)\over \alpha_1(t)} = {M_2(t)\over \alpha_2(t)} = 
{M_3(t)\over
\alpha_3(t)} = {M_{1/2}\over \alpha_G},  \label{Igauginogut}
\ee
where $\alpha_1(t_G) = \alpha_2(t_G) = \alpha_3(t_G) \equiv \alpha_G$
is the grand unified gauge coupling.
We note that the gaugino masses always satisfy the
relation~(\ref{Igauginogut}) 
irrespective of the breaking pattern~\cite{Kawamura1,Kawamura2, AP2} to the 
Standard Model gauge group so long as the underlying gauge group is
unified into a simple gauge group at a high mass scale $M_G$. 
We note that (\ref{Igauginogut})
is a result of one-loop renormalization group equations, and does not
hold at the two loop level~\cite{AP2}.
It follows from (\ref{Igauginoeq}) that
\be
M_i(t) = \alpha_i(t) {M_{\tilde g}\over \alpha_3({\tilde g})}.
\label{gauginorel}
\ee
Using the above, we can now express the functions $C_1$ and $C_2$
in terms of the gluino mass and the corresponding gauge couplings.
We have\cite{MR}
\bea
C_1(t) & = & {2 \over 11}{M_{\tilde g}^2\over \alpha_3^2(t_{\tilde g})}
[\alpha_1^2(t_{\tilde g}) - \alpha_1^2(t)], \label{Nc1eqn}\\
C_2(t) & = & {3 \over 2}{M_{\tilde g}^2\over \alpha_3^2(t_{\tilde g})}
[\alpha_2^2(t_{\tilde g}) - \alpha_2^2(t)]. \label{Nc2eqn}
\eea
We note that the gluino mass in (\ref{Nc3eqn}),\,(\ref{Nc1eqn})
and (\ref{Nc2eqn}) is the one-loop gluino mass and not the pole
mass, although these are related. 
Using these results for $C_i$, we can write the sum rules (\ref{Isum1})
and (\ref{Isum2}) as follows:
\bea
2 M_{\tilde{Q}}^2 - \msuR -\mseR &=& 
{M_{\tilde g}^2\over \alpha_3^2(t_{\tilde g})}
[{8\over 9} \alpha_3^2(t) - 3  \alpha_2^2(t) + {25 \over 99}\alpha_1^2(t)
+ {184 \over 99} \alpha^2_G], \label{sum1new}\\
M_{\tilde{Q}}^2 + \msdR - \mseR - M_{\tilde{L}}^2
&=& 
{M_{\tilde g}^2\over \alpha_3^2(t_{\tilde g})}
[{16\over 9} \alpha_3^2(t) + {20 \over 99}\alpha_1^2(t)
- {196 \over 99} \alpha^2_G].
\label{sum2new}
\eea
As before,
using a supersymmetric threshold of
$1$~TeV, and the values $M_G = 1.9 \times 10^{16}$ GeV, $\alpha_G = 0.04$,  
$\alpha_1(1~TeV) = 0.0173, \alpha_2(1~TeV) = 0.0328, \alpha_3(1~TeV) = 0.091$,
we can finally write our sum rules in terms of 
experimentally measurable masses as~(at a scale of $1$~TeV)
\bea
2 M_{\tilde{Q}}^2 - \msuR -\mseR &\simeq & 0.9 M_{\tilde g}^2,
\label{Isum1fin}\\
M_{\tilde{Q}}^2 + \msdR - \mseR - M_{\tilde{L}}^2 & \simeq &
1.4 M_{\tilde g}^2. \label{Isum2fin}
\eea
These relations, which relate the sfermion masses to the gluon mass,
are valid in $SU(5)$ as well as $SO(10)$ supersymmetric theories.

\newpage

\section{Sparticle Spectroscopy with Nonuniversal Gaugino Masses}
\label{nonuniv}
In the previous section we have considered the sparticle spectrum
in supersymmetric grand unified theories when the gaugino masses
are universal at the GUT scale. We have shown how the parameters
of the underlying supersymmetric grand unified theory  can be determined
from a measurement of sparticle masses.
However, universal gaugino masses are a very special case that can
arise in grand unified theories. In this section we shall explore
the changes in the sparticle spectrum  when the gaugino masses 
are non-universal at the GUT scale in the context of
$SU(5)$ and $SO(10)$ supersymmetric grand unified theories.

\subsection{Non-singlet chiral superfields}
In grand unified supersymmetric models
non-universal gaugino masses are generated by a 
non-singlet chiral superfield $\Phi^n$ that appears linearly in the gauge 
kinetic function $f(\Phi)$,  
which is an analytic function of the 
chiral superfields $\Phi$ in the theory.
The chiral superfields $\Phi$  are classified
into a set of gauge singlet superfields $\Phi^s$, and gauge nonsinglet
superfields $\Phi^n$, respectively under the grand unified group. 
If the auxiliary part $F_\Phi$ of a chiral superfield $\Phi$  in 
$f(\Phi)$ gets a VEV, then gaugino masses arise from the coupling of 
$f(\Phi)$ with the field strength superfield $W^a$. The Lagrangian
for the coupling of gauge kinetic function to the gauge field strength
is written as
\bea 
{\cal L}_{g.k.} \; = \;
\int d^2\theta f_{ab}(\Phi) W^{a}W^{b}
+h.c.,
\label{gk}
\eea
where $a$ and $b$ are gauge group indices, and repeated indices are summed 
over. The gauge kinetic function $f_{ab}(\Phi)$ is 
\bea
f_{ab}(\Phi) & = & f_0(\Phi^s)\delta_{ab} 
+ \sum_n f_n(\Phi^s){\Phi_{ab}^n\over M_P} + \cdot \cdot \cdot, 
\eea
where as indicated above the $\Phi^s$ and the  $\Phi^n$ are the singlet and 
the non-singlet chiral superfields, respectively. Here
$f_0(\Phi^s)$ and $f_n(\Phi^s)$ are functions of gauge singlet
superfields $\Phi^s$, and $M_P$ is some large scale.  When $F_\Phi$ 
gets a VEV $\langle F_\Phi \rangle$, the interaction~(\ref{gk})
gives rise to gaugino masses:  
\bea
{\cal L}_{g.k.} \; \supset \;
{{{\langle F_\Phi \rangle}_{ab}} \over {M_P}}
\lambda^a \lambda^b +h.c., 
\eea
where $\lambda^{a,b}$ are gaugino fields. Note that we denote by
$\lambda_1$, $\lambda_2$ and $\lambda_3$ the $U(1)$, $SU(2)$ and $SU(3)$ 
gauginos, respectively. Since the gauginos belong to the adjoint 
representation of $SU(5)$, $\Phi$ and $F_\Phi$ can belong to any of the 
following representations appearing in the symmetric product of the 
two {\bf 24} dimensional representations of $SU(5)$: 
\bea 
({\bf 24 \otimes 24})_{Symm} = {\bf 1 \oplus 24 \oplus 75 \oplus 200}.
\label{product}
\eea 
On the other hand the corresponding symmetric product
of the two { \bf 45} dimensional representations of
$SO(10)$ grand unified theory has the following decomposition:
\bea 
({\bf 45 \otimes 45})_{Symm} = {\bf 1 \oplus 54 \oplus 210\oplus 770}.
\label{product2}
\eea 


In the minimal, and the simplest, case $\Phi$ and $F_\Phi$ are
assumed to be in the singlet representation of 
$SU(5)$~(or $SO(10)$), which implies
equal gaugino masses at the GUT scale.  However, as is clear from the
decomposition (\ref{product}), $\Phi$
can belong to any of the non-singlet representations 
{\bf 24}, {\bf 75},  and {\bf 200} of $SU(5)$, in which case
these gaugino masses are unequal but related to one another via the
representation invariants~\cite{EENT}.  In Table~\ref{tab1}             
we show the ratios of resulting gaugino masses at tree-level as they
arise when $F_{\Phi}$ belongs to various representations of $SU(5)$,
and tabulate $a$ and $b$ defined so that at the unification scale
we have the ratio 
\be \label{ratios}
M_1:M_2:M_3 = a:b:1.
\ee

\begin{table}[b]
  \centering
\begin{tabular}{ccccccc} \hline 
    $F_\Phi$ & & & $a$ & & & $b$ \\ \hline
    {\bf 1} & & & 1 & & & 1 \\
    {\bf 24} & & & $-1/2$ & & & $-3/2$ \\ 
    {\bf 75} & & & $-5$ & & & 3 \\
    {\bf 200} & & & 10 & & & 2 \\ \hline 
  \end{tabular}
  \caption{The constants $a$ and $b$ for the various
representations of SU(5).}
\label{tab1}
\end{table}

Similarly, in Table~\ref{tab2} we show the corresponding $a$ and $b$ for
the $SO(10)$ grand unified theory. 
For definiteness, we shall study the case of each representation
separately, although an arbitrary combination of these is 
also allowed.~\footnote{ It has been pointed out~\cite{Andersonetal} 
that non-universal scalar masses can also arise from non-singlet 
chiral superfields, which are subdominant due to inverse powers of 
$M_G$ and read:
\begin{center}
$\displaystyle
{\cal L} \propto {\langle F_\Phi F_\Phi \rangle_{ij} 
\over M_G^2} \phi^\dag_i \phi_j, $ 
\end{center}
where the $\phi_{i,j}$ are the scalar fields.  
Recalling that for SU(5),
the left-handed SM fermions and their superpartners lie in
$\overline{{\bf 5}}-$ and ${\bf 10}$-dimensional representations,  while
the light Higgs bosons lie in the ${\bf 5}$ and $\overline{{\bf 5}}$, 
we need to consider the tensor products ${\bf 10}\otimes 
\overline{{\bf 10}} = {\bf 1}\oplus {\bf 24} \oplus {\bf 75}$ 
and ${\bf 5}\otimes \overline{{\bf 5}}={\bf 1}\oplus {\bf 24}$, and
deduce that for $\phi$ lying in ${\bf 1}$ and ${\bf 24}$
non-universality can be generated by this mechanism.
For  SO(10) with SM fermions lying in the ${\bf 16}$- and the 
light Higgs bosons in the ${\bf 10}$-dimensional representation, 
and the tensor decompositions ${\bf 16}\otimes {\bf 16}={\bf 1} \oplus
{\bf 45} \oplus {\bf 210}$, and ${\bf 10}\otimes {\bf 10}={\bf 1}_s\oplus
{\bf 45}_a\oplus {\bf 54}_s$, we conclude that the only instance when 
non-universality can be generated through this mechanism
is when $\phi$ lies in the singlet representation. We do not consider
this type of non-universality any further in this paper.}

\begin{table}[t] 
\centering
\begin{tabular}{ccccccccc} \hline
Case & & $F_\Phi$ & &Intermediate Stage&$a$ & & $b$ \\ \hline
A & & {\bf 1}  & &                    & 1 & & 1 \\
B & & {\bf 54} & & $G_{422}$          & $-1$ & & $-3/2$  \\
C & & {\bf 54} & &  $SU(2)\times SO(7)$ & 1 & & $-7/3$ \\
D & & {\bf 210}&& $H_{51}$           & $-96/25$ & & 1 \\ \hline
\end{tabular}
\caption{The constants $a$ and $b$ for $SO(10)$. Different cases
are labelled as  A, B, C and D,
in the notation of ref.~\cite{CHLW}.}
\label{tab2}
\end{table}

\subsection{Sum Rules}
We are now in a position to write down interrelationships
between the squarks and sfermions for the case of
non-universal gaugino masses.  Recalling from the case of universal
gaugino masses that 
\be
{M_i(t)\over \alpha_i(t)} = {M_i(t_G)\over \alpha_i(t_G)},
\label{gauginoeq}    
\ee
we have for the non-universal case
\be
{1\over a}{M_1(t)\over \alpha_1(t)} = {1\over b}{M_2(t)\over \alpha_2(t)} = 
{M_3(t)\over
\alpha_3(t)} = {M_{1/2}\over \alpha_G},  \label{gauginogut}
\ee
where $\alpha_1(t_G) = \alpha_2(t_G) = \alpha_3(t_G) \equiv \alpha_G$
is the grand unified gauge coupling, and  $a,\, b$ are as 
in (\ref{ratios}). Using this, the sum rules (\ref{sum1new}) and
(\ref{sum2new}) are now generalized for the case of 
non-universal gaugino masses to:
\bea
2 M_{\tilde{Q}}^2 - \msuR -\mseR &=& 
{M_{\tilde g}^2\over \alpha_3^2(t_{\tilde g})}
[{8\over 9} \alpha_3^2(t) - 3 b^2 \alpha_2^2(t) + {25 a^2\over 99}\alpha_1^2(t)
+ {(297 b^2-25 a^2-88) \over 99} \alpha^2_G], \label{Nsum1new}\\
M_{\tilde{Q}}^2 + \msdR - \mseR - M_{\tilde{L}}^2
&=& 
{M_{\tilde g}^2\over \alpha_3^2(t_{\tilde g})}
[{16\over 9} \alpha_3^2(t) + {20 a^2\over 99}\alpha_1^2(t)
- {4(5 a^2+44) \over 99} \alpha^2_G].
\label{Nsum2new}
\eea
Using the values for various couplings, as in the universal gaugino mass case,
these sum rules can finally be written as 
\bea
2 M_{\tilde{Q}}^2 - \msuR -\mseR &\simeq & K_1 M_{\tilde g}^2,
\label{sum1fin}\\
M_{\tilde{Q}}^2 + \msdR - \mseR - M_{\tilde{L}}^2 & \simeq &
K_2 M_{\tilde g}^2, \label{sum2fin}
\eea
where $K_1$ and $K_2$ are given in Table~\ref{tab3} and Table~\ref{tab4}
for $SU(5)$ and $SO(10)$, respectively.

\begin{table}[htb]
  \centering
\begin{tabular}{ccccccc} \hline 
    $F_\Phi$ & && $K_1$ &&& $K_2$ \\ \hline
    {\bf 1} &&& 0.9 &&& 1.4 \\
    {\bf 24} &&& $1.1$ &&& $1.4$ \\ 
    {\bf 75} &&& $1.4$ &&& $0.6$ \\
    {\bf 200} &&& $-2.5$&&& $-1.7$ \\ \hline 
  \end{tabular}
  \caption{The quantities  $K_1$ and $K_2$ for different 
  representations of SU(5).}
  \label{tab3}
\end{table}

\begin{table}[htb]  
\centering
\begin{tabular}{ccccccc} \hline
Case &&& $K_1$ &&& $K_2$ \\ \hline
A    &&                           & 0.9 &&& 1.4 \\
B    &&                            & $1.1$ &&& $1.4$  \\
C    &&                            & $1.7$ &&& $1.4$ \\
D    &&                            & $0.3$ &&& 1.0 \\ \hline
\end{tabular}
\caption{The quantities $K_1$ and $K_2$ for different
representations of $SO(10)$ labelled as A, B, C and D, see Table~\ref{tab2}.}
\label{tab4}
\end{table}
We should point out here that the determination of the parameters 
of the scalar sector as described in the previous Section can be carried
out in the case of non-universal gaugino masses 
with the replacements $\cone \to a^2 \cone,\, \ctwo \to b^2 \ctwo$,
with $a$ and $b$ given in Table~\ref{tab1} and Table~\ref{tab2}.

\section{Neutralinos and Charginos}
\label{neutchar}
As seen in the previous section, 
in supersymmetric theories with an underlying grand unified gauge group,
the gaugino masses need not be equal at the GUT scale. 
This may have important consequences for the neutralino 
and chargino mass spectrum.  We recall that neutralino is a much
favored  candidate for the lightest supersymmetric particle (LSP).
Thus, its composition is of crucial importance for supersymmetric 
phenomenology. In this section we shall consider some 
aspects of neutralino and chargino phenomenology when the
gaugino masses are non-universal at the grand unified scale.

We start by recalling the neutralino mass matrix in supersymmetric
models in the basis 
\bea \psi^0_j = (-i\lambda',~ -i\lambda_3,~
\psi^1_{H_1},~ \psi^2_{H_2}),~~~ j = 1, ~2, ~3, ~4,
\label{neut1}
\eea 
which can be written as~\cite{Nilles} 
\bea  
{\mathcal M} = 
\left( \begin{array}{cccc} M_1 & 0 & -M_Z\cos\beta\sin\theta_W &
M_Z\sin\beta\sin\theta_W\\ 0 & M_2 & M_Z\cos\beta\cos\theta_W & 
-M_Z\sin\beta\cos\theta_W\\
-M_Z\cos\beta\sin\theta_W& M_Z\cos\beta\cos\theta_W & 0 &-\mu\\
M_Z\sin\beta\sin\theta_W & -M_Z\sin\beta\cos\theta_W & -\mu & 0\\
\end{array} \right), \label{neutmatrix}
\eea 
where $\lambda'$ and $\lambda_3$ are the two-component gaugino
states corresponding to the $U(1)_Y$ and the third component of the
$SU(2)_L$ gauge groups, respectively, and $\psi^1_{H_1}, \psi^2_{H_2}$
are the two-component Higgsino states corresponding to the
two Higgs superfields $H_1$ and $H_2$ of the MSSM.  
We shall denote the eigenstates of the neutralino
mass matrix by $\chi^0_1, ~ \chi^0_2, ~ \chi^0_3, ~ \chi^0_4$ labeled in
order of increasing mass. Since some of the neutralino masses
resulting from diagonalization of the mass matrix can be negative, we
shall consider the squared mass matrix ${\mathcal
M}^{\dagger} {\mathcal M}$. 
Note that the masses and the compositions of neutralinos are
determined by the soft supersymmetry breaking gaugino masses $M_1$,
and $M_2$, the supersymmetric Higgs mixing parameter $\mu$,
and the ratio of the vacuum expectation values of the two
neutral Higgs bosons $H_1^0$ and $H_2^0$, $\langle H_2^0\rangle
/\langle H_1^0\rangle = \tan\beta$.

We now briefly consider the question of the composition of the
lightest neutralino with non-universal gaugino masses at the GUT scale
which follows from the neutralino mass matrix. 
Using the fact that the ratio $\alpha_2/\alpha_1$ is $\sim 2$ at the scale of
$1$ TeV, we conclude that for the case of $SU(5)$, with
$F_\Phi$ lying in ${\bf 1}$ and ${\bf 24}$, the ratio
$M_2\gg M_1$, and hence the lightest neutralino is mostly
bino-like~\cite{HLPR}.  On the other hand, with $F_\Phi$
lying in the ${\bf 75}$-dimensional representations, there are several 
possibilites depending on the values of the parameters. 
The lightest neutralino can be a bino for small values of $M_2$,
a wino for slightly larger values of  $M_2$, and a higgsino
for $M_2 \gsim 300$ GeV for values of $\tan\beta \gsim 10$.
Finally, for the ${\bf 200}$-dimensional representation
of $SU(5)$, the lightest neutralino can be either
a wino of a higgsino, depending on the values of
$M_2$ and $\tan\beta$.

For supersymemtric SO(10) grand unified theory, at a scale of $1$ TeV,
$M_2\gg M_1$ for models $A, \ B, \ C$, and hence 
the lightest neutralino is bino-like.  For model D, the magnitude of
$|a/b|\simeq 4$ and hence the lightest neutralino is  no longer a
bino-like state.  

It is useful to evaluate the upper bound on the mass of the lightest 
neutralino and the mass of the next to lightest neutralino in the 
grand unified models.  
We recall that a bound on the mass of the lightest neutralino
can be obtained by using the fact that the smallest eigenvalue of
${\mathcal M}^{\dagger} {\mathcal M}$ is smaller than the smallest 
eigenvalue of its upper left $2 \times 2$ submatrix, thereby
resulting in the upper bound~\cite{Pandita:1994zp, Pandita:1997zt} 
\begin{equation}
M^2_{\chi_1^0} \leq {1\over 2}[M_1^2+M_2^2+M_Z^2-
\sqrt{(M_1^2-M_2^2)+M_Z^4-2(M_1^2-M_2^2) M_Z^2 \cos 2\theta_W}].
\end{equation}
Similarly, one can write down the upper bound on the mass of the
second lightest neutralino from the structure of the neutralino
mass matrix~\cite{HLP}.

\begin{figure}[htb]
\psfrag{m}[r,t][r,t]{\rotatebox{90}{$m_{\tilde{\chi}_1^0}$}}
\psfrag{M3}[l,t][l,t]{$M_{\tilde{g}}\ \ \ $}
\includegraphics[width=6cm]{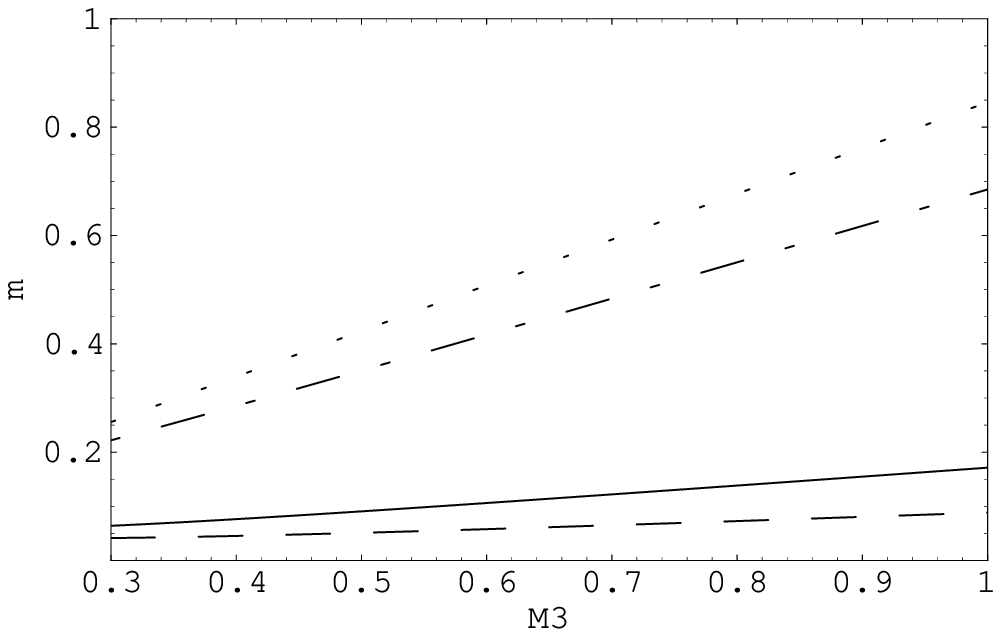} \hskip 1cm
\psfrag{m}[r,t][r,t]{\rotatebox{90}{$m_{\tilde{\chi}_2^0}$}}
\psfrag{M3}[l,t][l,t]{$M_{\tilde{g}}\ \ \ $}
\includegraphics[width=6cm]{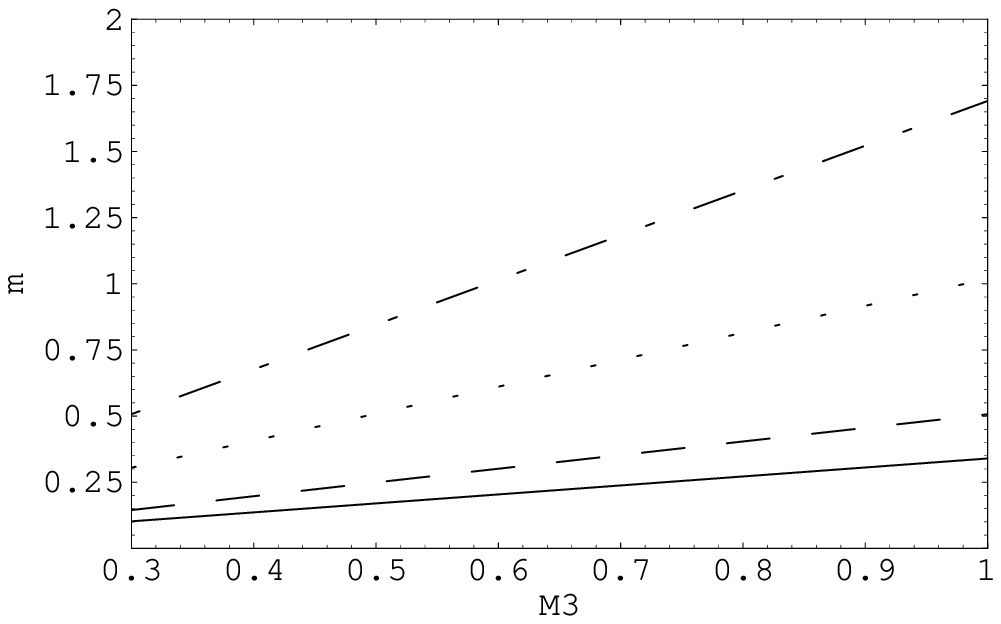} \hskip 1cm
\caption{Upper bound on the mass of the lightest neutralino and that of 
the next to lightest
neutralino as a function of the gluino mass for the four different cases 
that arise in $SU(5)$ 
supersymmetric grand unified theory.  The solid line
corresponds to {\bf 1}, the dashed to {\bf 24}, the dotted to {\bf 75} and
the dashed-dotted to {\bf 200} representations of $SU(5)$, 
respectively.  All masses are in TeV.}
\label{fig4}
\end{figure}

\begin{figure}[htb]
\psfrag{m}[r,t][r,t]{\rotatebox{+90}{$m_{\tilde{\chi}_1^0}$}}
\psfrag{M3}[l,t][l,t]{$M_{\tilde{g}}\ \ \ $}
\includegraphics[width=6cm]{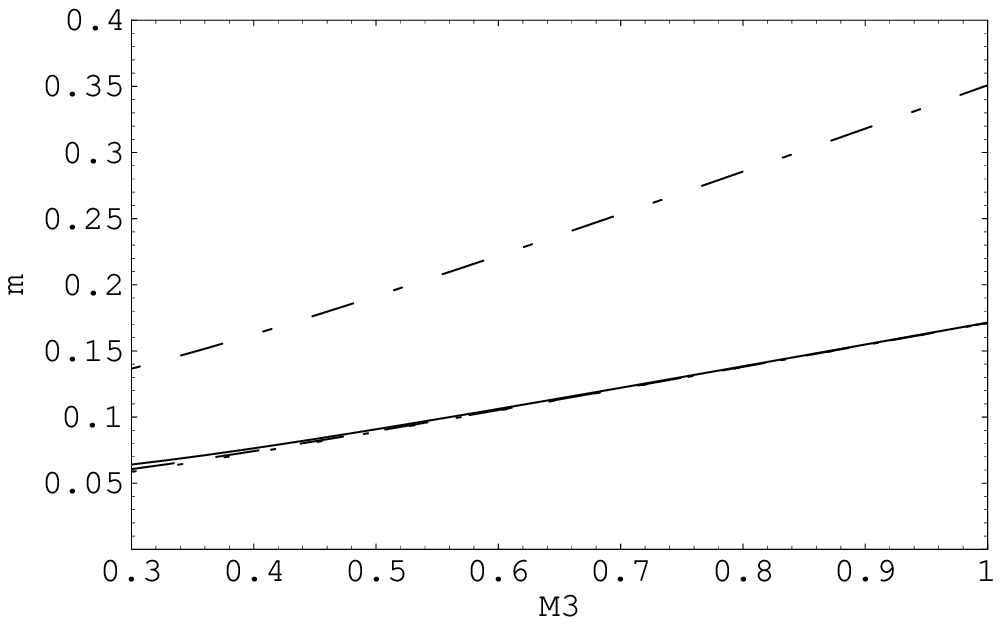} \hskip 1cm
\psfrag{m}[r,t][r,t]{\rotatebox{90}{$m_{\tilde{\chi}_2^0}$}}
\psfrag{M3}[l,t][l,t]{$M_{\tilde{g}}\ \ \ $}
\includegraphics[width=6cm]{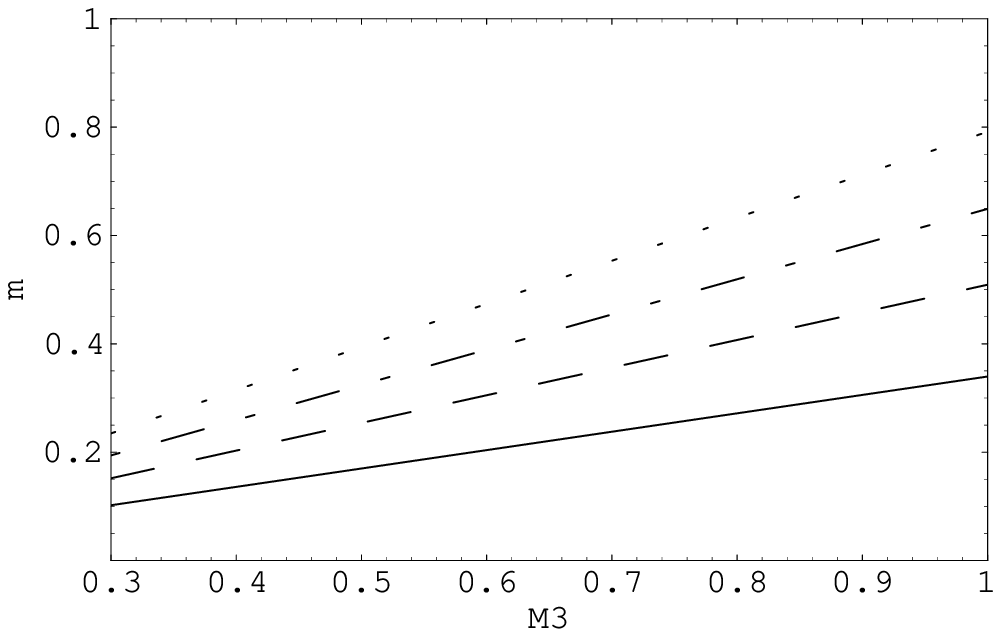} \hskip 1cm
\caption{Upper bound on the mass of the lightest neutralino and
that of the second lightest neutralino as a function of the gluino mass
for the SO(10) model.  The solid line
corresponds to A, the dashed to B, the dotted to C and
the dashed-dotted to D, respectively. The notation is as in 
Table~\ref{tab2}. All masses are in TeV.}
\label{fig5}
\end{figure}

In Fig.~\ref{fig4} and  Fig.~\ref{fig5}, we plot the upper bounds
on the mass of the lightest and the second lightest neutralino for the 
$SU(5)$ and $SO(10)$ supersymmetric grand unified theories,
respectively.  For the case of $SO(10)$ model, we see
that the bounds on the mass of the lightest neutralino for 
models A, B, C referred to in Table~\ref{tab2}
are numerically very similar. The upper bound in the case of model 
D is significantly larger
due to the numerical factor -96/25 that is present in the gaugino
mass ratio at the unification scale.  The bounds on the mass of the
next to lightest neutralino are significantly different for the 
different models as can be seen from the second panel in Fig.~\ref{fig5}.

\begin{figure}[htb]
\psfrag{m}[r,t][r,t]{\rotatebox{+90}{$M^2_{sum}$}}
\psfrag{M3}[l,t][l,t]{$M_{\tilde{g}}\ \ \ $}
\includegraphics[width=6cm]{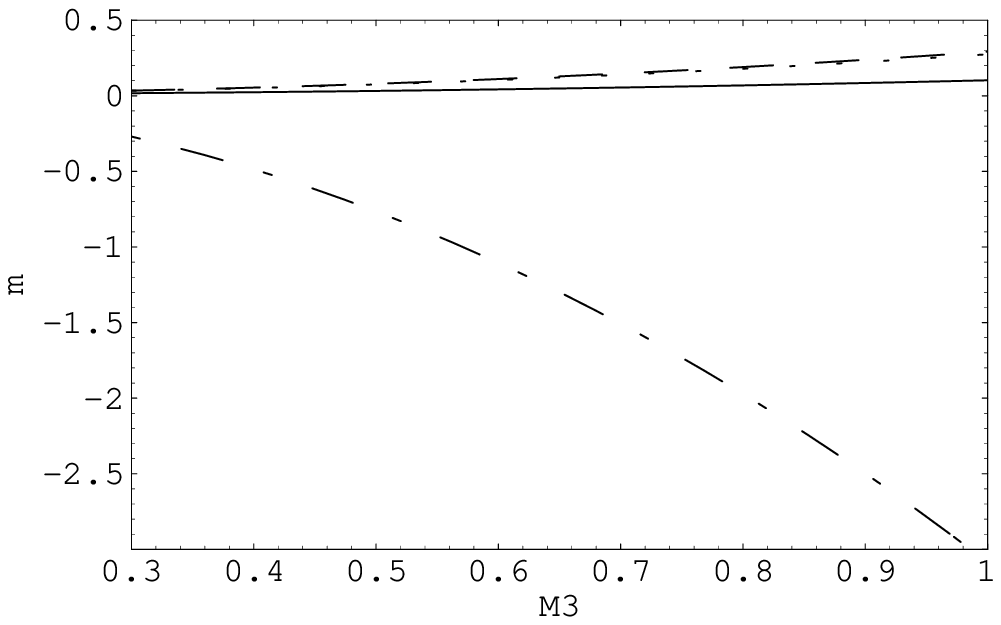} \hskip 1cm
\psfrag{m}[r,t][r,t]{\rotatebox{90}{$M^2_{sum}$}}
\psfrag{M3}[l,t][l,t]{$M_{\tilde{g}}\ \ \ $}
\includegraphics[width=6cm]{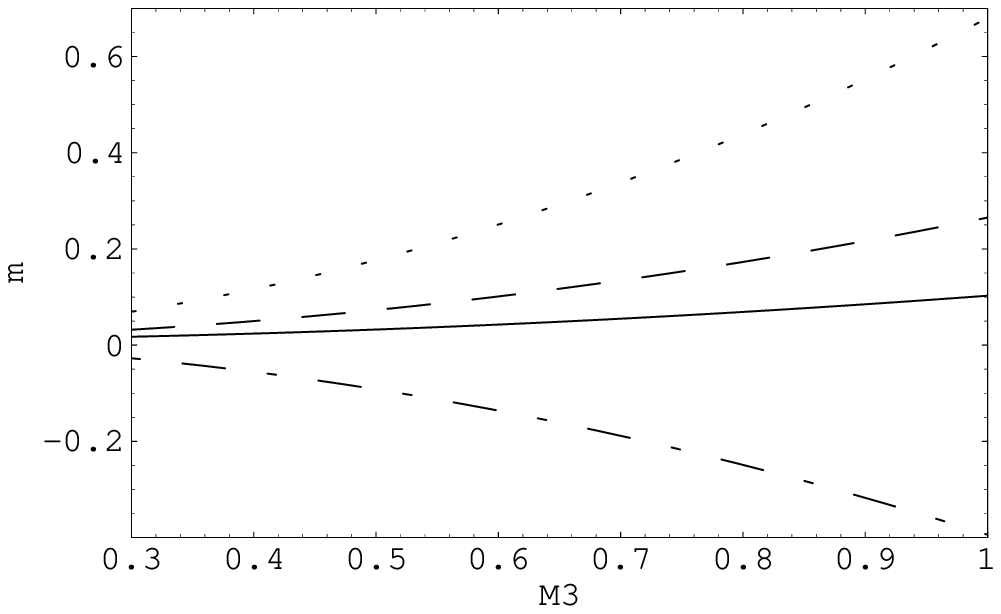} \hskip 1cm
\caption{The value of $M^2_{sum}$ as a function
of the gluino mass. In the  first panel 
we show this function for SU(5) models. In this panel solid line
corresponds to {\bf 1}, the dashed to {\bf 24}, the dotted to {\bf 75} and
the dashed-dotted to {\bf 200} dimensional representations
of $SU(5)$, respectively. In the second panel we plot this 
function for the SO(10) model, where the solid line
corresponds to A, the dashed to B, the dotted to C and
the dashed-dotted to D. The notation is as in Table~\ref{tab2}.  
All masses are in TeV.}
\label{fig6}
\end{figure}

The chargino mass matrix in  the basis $(-i\lambda^+, \psi^2_{H_1})$
can be written as 
\begin{equation}
\left(
\begin{array}{c c}
M_2 & \sqrt{2} M_W \sin\beta \\
\sqrt{2} M_W \cos\beta & \mu \\
\end{array}
\right),
\end{equation}
where $\lambda^+ = (1/\sqrt 2)(\lambda^1 - \lambda^2)$,
with $\lambda^1$ and $\lambda^2$ being the first and 
second component of $SU(2)_L$ gaugino, and 
$ \psi^2_{H_1}$ is the charged fermionic component of the 
$H_1$ superfield.
Taking the trace of the neutralino and chargino squared mass
matrices, we find the sum rule
\bea
M^2_{sum} = 2(M^2_{\tilde{\chi}_1^\pm}+M^2_{\tilde{\chi}_2^\pm})&-&
(M^2_{\tilde{\chi}_1^0}+M^2_{\tilde{\chi}_2^0}+M^2_{\tilde{\chi}_3^0}+
M^2_{\tilde{\chi}_4^0})\nonumber\\
&&=(b^2 \alpha_2^2- a^2 \alpha_1^2)\frac{M_{\tilde g}^2}{\alpha_3^2} +
4 m_W^2 -2m_Z^2. \label{neutsum}
\eea
Using the values of parameters $a$ and $b$ in Tables~\ref{tab1} 
and \ref{tab2}, the coefficient
of $M_{\tilde g}^2$ in (\ref{neutsum}) can be calculated
for various representations of $SU(5)$ and $SO(10)$ grand unified
theories. For $SU(5)$ the coefficient is $0.01,\, 0.28,\, 0.26$ 
and $-3.09$ 
for ${\bf 1,\, 24,\, 75}$ and ${\bf 200}$ representations, respectively,
whereas for $SO(10)$ this coefficient is evaluated to be
$0.01, \ 0.26, \ 0.67 $ and $-0.40$, respectively for the models 
$A, B, C, D$ in Table~\ref{tab2}. 
In Fig.~\ref{fig6} we plot $M^2_{sum}$ 
as a function of the gluino mass for $SU(5)$ and $SO(10)$
models. Thus, once the neutralino, chargino 
and gluino masses are measued experimentally, this sum rule can be used 
to distinguish between the gaugino non-universality that arises
in different supersymmetric grand unified theories.

\section{Sum rules for the third generation}
\label{thirdgen}
In section~\ref{sparticle} we have shown how a measurement of
sparticle masses of the light generations 
can be used to determine the parameters of the underlying 
supersymmetric theory. We also obtained relationships between
squarks and sleptons of the light generations
for different supersymmetric grand unified models.
We made use of of the explicit solutions
of the RG equations to arrive at these results. It is worthwhile 
to ask whether similar relations can be obtained for the 
squark and slepton masses of the third generation. In this section we
discuss this issue.

The renormlization group equations for the soft mass parameters
of squarks and sleptons of third generation
involve the Yukawa couplings of the third generation. These
are catalogued in Appendix~\ref{rgeall}. Because of the
dependence on third generation Yukawa
couplings, these equations cannot be solved analytically, unlike the
corresponding equations for the light generations.
However, the solutions for the third generation  RG equations can be 
written down in a closed form~\cite{Kazakov:1999pe}. These closed
form expressions  for the solutions of the third generation RG equations 
are written down in the Appendix~\ref{SOLRGE}. Note that we must add
the $D$-term contributions on the right hand side of the 
eqs.~(\ref{mass1}) - (\ref{mass5}) to get the masses of
the sfermions at the weak scale.  We first consider the case when 
the soft masses of the third generation are universal at the GUT scale:
\bea
m^2_{\tilde Q_{L3}}(t_G)  & = &  m^2_{\tilde t_R}(t_G) 
=  m^2_{\tilde b_R}(t_G) = m^2_{\tilde L_{L3}}  
=  m^2_{\tilde \tau_R}(t_G) = m_0^2,  \label{thirdsfcond0}\\
M^2_1(t_G) & = & M_2^2(t_G)= M_3^2(t_G)= M_{1/2}^2 \label{thirdgaugino0},
\eea
where now the values of the soft masses of sfermions refer 
to the third generation. As in the case of light generations,
we define the average values of the squark and slepton 
masses for the $SU(2)_L$ doublets of the third generation:
\bea
M_{\tilde{Q_3}}^2 & = & {1\over 2} (  M_{\tilde t_L}^2 +   M_{\tilde b_L}^2),  
\, \,  \, 
M_{\tilde{L_3}}^2 = {1\over 2} ( M_{\tilde \tau_L}^2  + 
M_{\tilde \nu_\tau}^2). \nonumber
\eea
Using the solutions of the RG equations for the third generation 
(\ref{mass1}) - (\ref{mass5}), as well as the solutions for the Higgs 
mass parameters (\ref{mass6}) - (\ref{mass7}), we obtain the mass
relation:
\bea
[M_{\tilde{Q_3}}^2 - 2  M_{\tilde t_R}^2 + 
M_{\tilde b_R}^2 +  M_{\tilde \tau_R}^2  - M_{\tilde{L_3}}^2]
+ (m_{H_2}^2 - m_{H_1}^2) & = & - \frac{10}{3} \sin^2\theta_W M_Z^2 \cos 2\beta.
\label{third1}
\eea
However, if the boundary conditions for the soft scalar masses are 
non-universal, as happens in $SU(5)$ and $SO(10)$ grand unified theories,
then the mass relation (\ref{third1}) is modified to
\bea
[M_{\tilde{Q_3}}^2 - 2  M_{\tilde t_R}^2 +
M_{\tilde b_R}^2 +  M_{\tilde \tau_R}^2  - M_{\tilde{L_3}}^2]
+ (m_{H_2}^2 - m_{H_1}^2)
& = & - \frac{10}{3} \sin^2\theta_W M_Z^2 \cos 2\beta
 \nonumber \\
&& + S(t_G)\left [\frac{20}{33} + \frac{13}{33} \frac {\alpha_1(t)}
{\alpha_1(t_G)}\right ].
\label{third2}
\eea
We note that the Higgs mass parameters enter the relations
(\ref{third1}) and (\ref{third2}) by virtue of the fact
that the evolution of soft masses for the third generation
is coupled to evolution of the Higgs mass parameters through the
large Yukawa couplings of the third generation.

\section{Summary and Discussion}
In this paper we have considered the spectrum of sparticles 
as it arises in $SU(5)$ and $SO(10)$ supersymmetric grand unified
theories. We have shown through an analytical solution of the
one-loop renormalization group equation how the measurement of 
the sparticle masses
can be used to obtain the values of parameters 
of an underlying supersymmetric theory. It may, thus, be possible
to distinguish, from the
reconstruction of these parameters, 
between different supersymmetric grand unified theories.
We have emphasized that 
non-universality of the soft scalar masses can arise because of
an underlying grand unified theory. We have also discussed
the non-universality that can arise in the gaugino sector,
and its implications for the sparticle spectrum.
In particular, we have considered different interrelationships
between sfermion masses and the gluino mass that can arise
in supersymmetric grand unified theories. We have also considered
the implications of an underlying grand unified theory
for the neutralino and chargino spectrum, which may be 
of direct relevance for  sparticle  searches at the LHC.

Our analysis has been based on the solutions of the one-loop renormalization
group equations supplemented by the boundary conditions flowing from
the underlying grand unified gauge group. Thus, the grand unified
symmetry leaves its imprint on the superparticle spectra through
these boundary conditions.  However, there may be significant 
corrections to our results. These can arise from higher loop 
contributions to the renormalization group equations, use of 
a particular renormalization scheme as well as theoretical uncertainties.
Since we  have considered boundary conditions arising from different 
grand unified theories, GUT scale threshold can also affect our results.
In testing our results, there will also uncertainties arising
from sparticle masses, which will propagate to the underlying parameters.
We have not considered these corrections in our paper. Our aim has been to
show through transparent analytic calculations how different grand 
unified theories can be tested through a maesurement of sparticle masses.  
Corrections to our results can be calculated only through a detailed
analysis. We hope to calculate these corrections in a future publication.

\label{summary}

\section{Acknowledgements}
PNP wishes to thank the Institute of Mathematical Sciences, Chennai
for hospitality while this paper was completed.
BA thanks the Department of Science and
Technology, and Council for Scientific and Industrial Research,
India, whereas
PNP thanks the Department of Atomic Energy,  and Council for
Scientific and Industrial Research, India,
for support during the course of this work.
\begin{appendix}
\section{Renormalization Group Equations}
\label{rgeall}
The renormalization group equations for the soft supersymmetry
breaking mass parameters and the
Higgs mass parameters can be written as
\bea
16 \pi^2 {d  \over d t} m^2_{{\tilde t}_L, {\tilde b}_L} & = &
2 y_t^2 X^2_t + 2 y_b^2 X^2_b - {2 \over 15} g_1^2 M_1^2
- 6  g_2^2 M_2^2
-{32\over 3}  g_3^2 M_3^2
+ {1\over 5} g_1^2 S, \label{rg1} \\
16 \pi^2 {d  \over d t} m^2_{{\tilde t}_R} & = &
4 y_t^2 X^2_t -{32 \over 15} g_1^2 M_1^2
-{32\over 3}  g_3^2 M_3^2
- {4\over 5} g_1^2 S,  \label{rg2} \\
16 \pi^2 {d  \over d t} m^2_{{\tilde b}_R} & = &
4 y_b^2 X^2_b -{8 \over 15} g_1^2 M_1^2
-{32\over 3}  g_3^2 M_3^2
+ {2\over 5} g_1^2 S,  \label{rg3} \\ 
16 \pi^2 {d  \over d t} m^2_{ {\tilde \nu}_\tau, {\tilde \tau}_L} & = &
2 y_\tau^2 X^2_\tau  - {6 \over 5} g_1^2 M_1^2
- 6  g_2^2 M_2^2
- {3\over 5} g_1^2 S,  \label{rg4} \\
16 \pi^2 {d  \over d t} m^2_{{\tilde \tau}_R} & = &
4 y_\tau^2 X^2_\tau -{24 \over 5} g_1^2 M_1^2
+ {6\over 5} g_1^2 S,  \label{rg5}\\
16 \pi^2 {d  \over d t} m^2_{{\tilde N}_3} & = & 4 y_{\nu} X^2_{\nu},
\label{rg6} \\
16 \pi^2 {d  \over d t} m^2_{H_d}  & = & 6 y_b^2 X_b^2 
+ 2 y_{\tau}^2 X_{\tau}^2
- {6\over 5} g_1^2 M_1^2 -  6 g_2^2 M_2^2 - {3\over 5} g_1^2 S,  \label{rg7}\\
16 \pi^2 {d  \over d t} m^2_{H_u}  & = & 6 y_t^2 X_t^2
+ 2 y_{\tau}^2 X_{\tau}^2
- {6\over 5} g_1^2 M_1^2 -  6 g_2^2 M_2^2 + {3\over 5} g_1^2 S, \ \label{rg8}
\eea
where $t \equiv {\rm ln}(Q/M_G)$, with $M_G$ being the initial 
GUT scale, $y_i (i = t, b, \tau, \nu)$ are the Yukawa couplings,
$M_{3,2,1}$ are the running gaugino masses 
and $g_{3,2,1}$ are the
corresponding gauge couplings associated with the SM gauge 
group~($\alpha_i \equiv g_i^2/4\pi$, 
$g_1$ is in GUT normalization), and
\bea
X_t^2 & = &(m_{H_u}^2 + m_{{\tilde t}_L}^2 + m_{{\tilde t}_R}^2 + 
A_t^2 ),\\
X_b^2 & = &(m_{H_d}^2 + m_{{\tilde b}_L}^2 + m_{{\tilde b}_R}^2 + 
A_b^2 ), \\
X_\tau^2 & = &(m_{H_d}^2 + m_{{\tilde \tau}_L}^2 
+ m_{{\tilde \tau}_R}^2 + A_\tau^2),  \\
X_{\nu}^2 & = &  (m_{H_u}^2 +  m_{{\tilde \nu}_{\tau}}^2 
+  m_{{\tilde N}_3}^2 + A_{\nu}^2), \\
S & \equiv & {\rm Tr}(Ym^2) =  m_{H_u}^2 - m_{H_d}^2 + \sum_{\rm families}
(\msQ - 2 \msuR + \msdR - \mslL + \mseR).
\eea
Here $A_k~(k = t, b, \tau,  \nu)$ are the  soft supersymmetry 
breaking trilinear couplings,  and $S$, with $Y$ being the hypercharge,
is the parameter which is a measure of  the nonuniversality of the soft 
scalar masses.  Note that we have included in the above the renormalization
group equation for the soft mass parameter for the right handed sneutrino.

The quantity S evolves according to~($\tilde\alpha_i \equiv g_i^2/16\pi^2$)
\be
{dS(t) \over dt} = {66\over 5} {\tilde\alpha_1}S(t),
\label{sequation1}
\ee
which has the solution
\bea
S(t) &  = & S(t_G) {\tilde \alpha_1(t) \over \tilde \alpha_1(t_G)}.
\label{sequation2}
\eea
We note that if $S = 0$ at some initial scale, which would be the case if
all the soft sfermion and Higgs masses are equal at that scale,
\bea
S & \equiv &  {\rm Tr}(Ym^2) = m_0^2 \ {\rm Tr}Y = 0, \label{sequation3}
\eea
then the RG evolution
will maintain it to be zero at all scales.  However, in a typical GUT,
like $SU(5)$ or $SO(10)$, this is not the case, and $S$ is nonzero at the
scale where the GUT gauge group is broken. In this paper, unless otherwise
stated, we shall take $S$ to be nozero.

The renormalization group equations for the gauge couplings, 
the gaugino masses, the Yukawa couplings, and the $A$ parameters 
can be written as
\bea
16 \pi^2  {d \over d t}g_i & = & -b_i g_i^3,  \label{gaugerg}\\
16 \pi^2  {d \over d t}M_i  & = & - 2 b_i M_i g_i^2,  \label{gauginorg}\\
{d \over d t}Y_k  & = &
Y_k(\sum_{l}a_{kl}Y_l - \sum_{i}c_{ki}\alpha_i ), \label{yukawarge}\\
{d \over d t} A_k & = &  \sum_{l}a_{kl}A_l - \sum_{i}c_{ki}\alpha_i M_i, 
\label{Arge}
\eea
where
\begin{eqnarray}
b_i & = &\{-33/5, -1, 3 \}, \\
c_{ti} & = & \{13/15,3,16/3 \}, \, \,
c_{bi}  =  \{7/15,3,1 6/3 \},  \\
c_{\tau i} & = &\{9/5,3,0 \}, \, \,
c_{\nu i}  =  \{3/5,3,0 \}, \\
a_{tl} & = &\{6,1,0,1 \}, \, \, 
a_{bl}  =  \{1,6,1,0 \}, \\
a_{\tau l} & = &\{0,3,4,1\}, \, \,
a_{\nu l}  = \{3,0,1,4\},
\end{eqnarray}

\section{Solution of Renormalization Group Equations}
\label{SOLRGE}
Here we present the solutions of the renormalization
group equations for the third generation given in
Appendix~\ref{rgeall} in a closed form.
\begin{eqnarray}
{m}^2_{{{\tilde t}_L, {\tilde b}_L}}
&=&\tilde{m}_{{{\tilde t}_L, {\tilde b}_L}}^2(t_G)
+\frac{48 C_3+ 58C_2 - 55/6 C_1}{122}
+\frac{17(\Sigma_t-\Sigma_t^0)+20(\Sigma_b-\Sigma_b^0)
-5(\Sigma_\tau-\Sigma_\tau^0)}{122} - \frac{1}{5} K,\label{mass1} \\
{m}^2_{{\tilde t}_R}
&=&\tilde{m}_{{\tilde t}_R}^2(t_G)+\frac{54 C_3- 72 C_2+ 24 C_1}{122}
+\frac{42(\Sigma_t-\Sigma_t^0)-8(\Sigma_b-\Sigma_b^0)
+2(\Sigma_\tau-\Sigma_\tau^0)}{122} + \frac{4}{5} K,\label{mass2}\\
{m}^2_{{\tilde b}_R}
&=&\tilde{m}_{{\tilde b}_R}^2(t_G)+\frac{42C_3- 56 C_2+112/6C_1}{122}
+\frac{-8(\Sigma_t-\Sigma_t^0)+48(\Sigma_b-\Sigma_b^0)
-12(\Sigma_\tau-\Sigma_\tau^0)}{122} - \frac{2}{5} K,\label{mass3}\\
{m}^2_{ {\tilde \nu}_\tau, {\tilde \tau}_L}
&=& \tilde{m}_{ {\tilde \nu}_\tau, {\tilde \tau}_L}^2(t_G)
+\frac{30C_3+82C_2-103/6C_1}{122}
+\frac{3(\Sigma_t-\Sigma_t^0)-18(\Sigma_b-\Sigma_b^0)
+35(\Sigma_\tau-\Sigma_\tau^0)}{122} + \frac{3}{5} K,\label{mass4}\\
{m}^2_{ {\tilde \tau}_R}
&=&\tilde{m}_{ {\tilde \tau}_R}^2(t_G)+\frac{60C_3-80C_2+ 80/3 C_1}{122}
+\frac{6(\Sigma_t-\Sigma_t^0)-36(\Sigma_b-\Sigma_b^0)
+70(\Sigma_\tau-\Sigma_\tau^0)}{122} - \frac{6}{5} K,\label{mass5}\\
m^2_{H_{d}}&=&{m}_{H_{d}}^2(t_G)+\frac{-90C_3-2C_2-57/6 C_1}{122}
+\frac{-9(\Sigma_t-\Sigma_t^0)+54(\Sigma_b-\Sigma_b^0)
+17(\Sigma_\tau-\Sigma_\tau^0)}{122} + \frac{3}{5} K ,\label{mass6}\\
{m}^2_{H_{u}}&=&{m}_{H_{u}}^2(t_G) +\frac{-102C_3+ 14C_2-89/6C_1}{122}
+\frac{63(\Sigma_t-\Sigma_t^0)-12(\Sigma_b-\Sigma_b^0)
+3(\Sigma_\tau-\Sigma_\tau^0)}{122} - \frac{3}{5} K,\label{mass7}
\end{eqnarray}
where $C_i$ are defined in (\ref{cidef}), and
\bea
\Sigma_t & = &(m_{H_u}^2 + m_{{\tilde t}_L}^2 + m_{{\tilde t}_R}^2),\\
\Sigma_b^2 & = &(m_{H_d}^2 + m_{{\tilde b}_L}^2 + m_{{\tilde b}_R}^2), \\
\Sigma_\tau^2 & = &(m_{H_d}^2 + m_{{\tilde \tau}_L}^2+ m_{{\tilde \tau}_R}^2),  \\
\Sigma_k^0 & = & \Sigma_k(t_G), ~~~~ k = t, b, \tau.
\eea
The values of $\Sigma_k$ completely
define those of the soft masses for squarks, sleptons and Higgs bosons
due to linear relations which follow from the RG 
equations~\cite{Kazakov:1999pe}.  Note that we have included here
the contribution comimg from the  nonunivesality of the soft masses 
through the parameter $K$ which was neglected in~\cite{Kazakov:1999pe}.

\end{appendix}


\begin{thebibliography}{abcdef}



\bibitem{GG1}
  H.~Georgi and S.~L.~Glashow,
  Phys.\ Rev.\ Lett.\  {\bf 32}, 438 (1974).

\bibitem{Seesaw1}
  P.~Minkowski,
  Phys.\ Lett.\ B {\bf 67}, 421 (1977).

\bibitem{Seesaw2}
  M.~Gell-Mann, P.~Ramond and R.~Slansky,
Print-80-0576 (CERN).

\bibitem{Seesaw3}
  T.~Yanagida,
{\it In Proceedings of the Workshop on the Baryon Number of the Universe 
and Unified Theories, Tsukuba, Japan, 13-14 Feb., 1979}.

\bibitem{Seesaw4}
  R.~N.~Mohapatra and G.~Senjanovic,
  Phys.\ Rev.\ Lett.\  {\bf 44}, 912 (1980).

\bibitem{SO101}
  H.~Georgi,
  AIP Conf.\ Proc.\  {\bf 23}, 575 (1975).

\bibitem{SO102}
  H.~Fritzsch and P.~Minkowski,
  Annals Phys.\  {\bf 93}, 193 (1975).

\bibitem{GQW}
  H.~Georgi, H.~R.~Quinn and S.~Weinberg,
  Phys.\ Rev.\ Lett.\  {\bf 33}, 451 (1974).

\bibitem{wess}
J.~Wess and J.~Bagger,
``Supersymmetry and supergravity,'' Princeton University Press~(1992)

\bibitem{kaul1}
G.~'t Hooft in {\it Recent Developments in  Gauge Theories},
Eds. G.~'t Hooft et al.~( Plenum Press, New York, 1980), p 438.
\bibitem{kaul2}
E.~Witten,
Nucl.\ Phys.\ B {\bf 188}, 513 (1981).
\bibitem{kaul3}
R.~K.~Kaul,
Phys.\ Lett.\ B {\bf 109}, 19 (1982).
\bibitem{kaul4}
R.~K.~Kaul and P.~Majumdar,
Nucl.\ Phys.\ B {\bf 199}, 36 (1982).
\bibitem{kaul5}
R.~K.~Kaul,
Pramana {\bf 19}, 183 (1982).

\bibitem{Nilles}
H.~P.~Nilles,
Phys.\ Rept.\  {\bf 110}, 1 (1984).

\bibitem{LSusyreview}
D.~J.~H.~Chung, L.~L.~Everett, G.~L.~Kane, S.~F.~King, J.~D.~Lykken 
and L.~T.~Wang,
Phys.\ Rept.\  {\bf 407}, 1 (2005)
[arXiv:hep-ph/0312378].

\bibitem{dimopoulos}
  S.~Dimopoulos, S.~Raby and F.~Wilczek,
      Phys.\ Rev.\ D {\bf 24}, 1681 (1981).



\bibitem{Lang1}
  P.~Langacker and N.~Polonsky,
  Phys.\ Rev.\ D {\bf 52}, 3081 (1995)
  [arXiv:hep-ph/9503214].
\bibitem{Lang2}
 U.~Amaldi, W.~de Boer and H.~Furstenau,
  Phys.\ Lett.\  B {\bf 260}, 447 (1991).
\bibitem{Lang3}
C.~Giunti, C.~W.~Kim and U.~W.~Lee,
  Mod.\ Phys.\ Lett.\  A {\bf 6}, 1745 (1991).



\bibitem{AFM}
  G.~Altarelli, F.~Feruglio and I.~Masina,
  JHEP {\bf 0011}, 040 (2000)
  [arXiv:hep-ph/0007254].

\bibitem{alfe}
  G.~Altarelli and F.~Feruglio,
  New J.\ Phys.\  {\bf 6}, 106 (2004)
  [arXiv:hep-ph/0405048].

\bibitem{Inoue:1982pi}
  K.~Inoue, A.~Kakuto, H.~Komatsu and S.~Takeshita,
  Prog.\ Theor.\ Phys.\  {\bf 68}, 927 (1982)
  [Erratum-ibid.\  {\bf 70}, 330 (1983)].

\bibitem{LHCILCReview}
  G.~Weiglein {\it et al.}  [LHC/LC Study Group],
  Phys.\ Rept.\  {\bf 426}, 47 (2006)
  [arXiv:hep-ph/0410364].

\bibitem{FHKN}
  A.~E.~Faraggi, J.~S.~Hagelin, S.~Kelley and D.~V.~Nanopoulos,
  Phys.\ Rev.\ D {\bf 45}, 3272 (1992).

\bibitem{BPZ}
  G.~A.~Blair, W.~Porod and P.~M.~Zerwas,
  Eur.\ Phys.\ J.\ C {\bf 27}, 263 (2003)
  [arXiv:hep-ph/0210058].


\bibitem{Blair}
  G.~A.~Blair, A.~Freitas, H.~U.~Martyn, G.~Polesello, W.~Porod and P.~M.~Zerwas,
  Acta Phys.\ Polon.\  B {\bf 36}, 3445 (2005)
  [arXiv:hep-ph/0512084].

\bibitem{Kane}
  G.~L.~Kane, P.~Kumar, D.~E.~Morrissey and M.~Toharia,
  arXiv:hep-ph/0612287.

\bibitem{Ramond:1979py}
  P.~Ramond,
  arXiv:hep-ph/9809459.

\bibitem{Drees:1986vd}
  M.~Drees,
  Phys.\ Lett.\ B {\bf 181}, 279 (1986).

\bibitem{ChengHall}
  H.~C.~Cheng and L.~J.~Hall,
            Phys.\ Rev.\ D {\bf 51}, 5289 (1995)
	            [arXiv:hep-ph/9411276].



\bibitem{Kolda:1995iw}
  C.~F.~Kolda and S.~P.~Martin,
  Phys.\ Rev.\ D {\bf 53}, 3871 (1996)
  [arXiv:hep-ph/9503445].

\bibitem{Auto:2003ys}
  D.~Auto, H.~Baer, C.~Balazs, A.~Belyaev, J.~Ferrandis and X.~Tata,
  JHEP {\bf 0306}, 023 (2003)
  [arXiv:hep-ph/0302155].

	  

\bibitem{Kawamura1}
  Y.~Kawamura, H.~Murayama and M.~Yamaguchi,
  Phys.\ Lett.\ B {\bf 324}, 52 (1994)
  [arXiv:hep-ph/9402254].

\bibitem{Kawamura2}
  Y.~Kawamura, H.~Murayama and M.~Yamaguchi,
  Phys.\ Rev.\ D {\bf 51}, 1337 (1995)
  [arXiv:hep-ph/9406245].

\bibitem{EENT}
  J.~R.~Ellis, K.~Enqvist, D.~V.~Nanopoulos and K.~Tamvakis,
  Phys.\ Lett.\ B {\bf 155}, 381 (1985).

\bibitem{DreesKE}
  M.~Drees,
  Phys.\ Lett.\ B {\bf 158}, 409 (1985).

\bibitem{coll1}
  V.~D.~Barger and C.~Kao,
  Phys.\ Rev.\ D {\bf 60}, 115015 (1999)
  [arXiv:hep-ph/9811489].

\bibitem{coll2}
  G.~Anderson, H.~Baer, C.~h.~Chen and X.~Tata,
  Phys.\ Rev.\ D {\bf 61}, 095005 (2000)
  [arXiv:hep-ph/9903370].

\bibitem{constr1}
  K.~Huitu, Y.~Kawamura, T.~Kobayashi and K.~Puolamaki,
  Phys.\ Rev.\ D {\bf 61}, 035001 (2000)
  [arXiv:hep-ph/9903528].

\bibitem{constr2}
  G.~Belanger, F.~Boudjema, A.~Cottrant, A.~Pukhov and A.~Semenov,
  Nucl.\ Phys.\ B {\bf 706}, 411 (2005)
  [arXiv:hep-ph/0407218].

\bibitem{DMM}
  A.~Djouadi, Y.~Mambrini and M.~Muhlleitner,
  Eur.\ Phys.\ J.\ C {\bf 20}, 563 (2001)
  [arXiv:hep-ph/0104115].

\bibitem{HLPR}
  K.~Huitu, J.~Laamanen, P.~N.~Pandita and S.~Roy,
  Phys.\ Rev.\ D {\bf 72}, 055013 (2005)
  [arXiv:hep-ph/0502100].

\bibitem{KLNPY}
  S.~Kelley, J.~L.~Lopez, D.~V.~Nanopoulos, H.~Pois and K.~j.~Yuan,
  Phys.\ Rev.\ D {\bf 47}, 2461 (1993)
  [arXiv:hep-ph/9207253].


\bibitem{hep-ph/0211071}
  A.~Birkedal-Hansen and B.~D.~Nelson,
  Phys.\ Rev.\ D {\bf 67}, 095006 (2003)
  [arXiv:hep-ph/0211071].

\bibitem{HLP}
  K.~Huitu, J.~Laamanen and P.~N.~Pandita,
  Phys.\ Rev.\ D {\bf 67}, 115009 (2003)
  [arXiv:hep-ph/0303262].
                                                                                
\bibitem{Profumo}
  S.~Profumo,
  Phys.\ Rev.\ D {\bf 68}, 015006 (2003)
  [arXiv:hep-ph/0304071].

\bibitem{Pati:1974yy}
  J.~C.~Pati and A.~Salam,
  Phys.\ Rev.\ D {\bf 10}, 275 (1974).

\bibitem{AP2}
  B.~Ananthanarayan and P.~N.~Pandita,
  Int.\ J.\ Mod.\ Phys.\ A {\bf 20}, 4241 (2005)
  [arXiv:hep-ph/0412125].

\bibitem{AP1}
  B.~Ananthanarayan and P.~N.~Pandita,
  Mod.\ Phys.\ Lett.\ A {\bf 19}, 467 (2004)
  [arXiv:hep-ph/0312361].

\bibitem{MR}
  S.~P.~Martin and P.~Ramond,
  Phys.\ Rev.\ D {\bf 48}, 5365 (1993)
  [arXiv:hep-ph/9306314].

\bibitem{Andersonetal}
G.~Anderson, C.~H.~Chen, J.~F.~Gunion, J.~D.~Lykken, T.~Moroi and Y.~Yamada,  
    eConf {\bf C960625}, SUP107 (1996)
     [arXiv:hep-ph/9609457].

\bibitem{CHLW}
  N.~Chamoun, C.~S.~Huang, C.~Liu and X.~H.~Wu,
      Nucl.\ Phys.\ B {\bf 624}, 81 (2002)
        [arXiv:hep-ph/0110332].

\bibitem{Pandita:1994zp}
  P.~N.~Pandita,
        Phys.\ Rev.\  D {\bf 53}, 566 (1996).

\bibitem{Pandita:1997zt}
 P.~N.~Pandita,
     arXiv:hep-ph/9701411.

\bibitem{Kazakov:1999pe}
  D.~Kazakov and G.~Moultaka,
  Nucl.\ Phys.\ B {\bf 577}, 121 (2000)
  [arXiv:hep-ph/9912271].

\end{thebibliography}
\end{document}